\documentclass[a4paper,10pt]{article}
\usepackage{url}
\usepackage{graphicx}				
\usepackage{amsmath}				
\usepackage{amsfonts}
\usepackage[colorlinks=true,linkcolor=black,anchorcolor=black,citecolor=blue,filecolor=red,menucolor=black,urlcolor=blue,breaklinks=true,pdfhighlight=/P,pdfmenubar=true,pdftoolbar=true,pdfpagelabels=true,pdfstartpage=1,pdfstartview=FitV,pdftitle={Quickest Paths in Simulations of Pedestrians},pdfsubject={Pedestrian Route Choice},pdfauthor={Kretz},pdfcreator={Kretz},pdfproducer={Kretz},pdfkeywords={pedestrian route choice simulation VISSIM assignment quickest path space time}]{hyperref}	
\usepackage[numbers,sort&compress]{natbib}	
\usepackage{hypernat}				
\usepackage[american]{babel}
\usepackage{simplemargins}
\setallmargins{1in}

\begin{document}

\markboth{T. Kretz, A. Gro{\ss}e, S. Hengst, L. Kautzsch, A. Pohlmann, and P. Vortisch}{Quickest Paths in Simulations of Pedestrians}
%
%

\title{Quickest Paths in Simulations of Pedestrians}

\author{Tobias Kretz$^{p}$, Andree Gro{\ss}e$^{p}$, Stefan Hengst$^{p}$, \\ Lukas Kautzsch$^{p}$, Andrej Pohlmann$^{p}$, Peter Vortisch$^{k}$ \\ \\
$^{p}$: PTV Planung Transport Verkehr AG \\
Stumpfstr. 1, \\ D-76131 Karlsruhe, Germany\\
{\tt\{forename.surname\}@ptv.de} \\ \\
$^{k}$: Institut f{\"u}r Verkehrswesen, \\ 
Karlsruhe Institute of Technology (KIT) \\ 
Otto Ammann-Platz 9 \\
D-76128 Karlsruhe, Germany \\ 
{\tt Peter.Vortisch@kit.edu}}

\maketitle

\newlength{\figurewidth}\setlength{\figurewidth}{0.618\textwidth}
\begin{abstract}%
This contribution proposes a method to make agents in a microscopic simulation of pedestrian traffic walk approximately along a path of estimated minimal remaining travel time to their destination. Usually models of pedestrian dynamics are (implicitly) built on the assumption that pedestrians walk along the shortest path. Model elements formulated to make pedestrians locally avoid collisions and intrusion into personal space do not produce motion on quickest paths. Therefore a special model element is needed, if one wants to model and simulate pedestrians for whom travel time matters most (e.g. travelers in a station hall who are late for a train). Here such a model element is proposed, discussed and used within the Social Force Model.
\end{abstract}


\section{Introduction}
\subsection{Motivation: Travel Time matters for Pedestrians}
In traffic planning for vehicular traffic it is common sense for more than half a century that -- provided origin-destination matrices are known -- traffic demand on the links of the network cannot be predicted based on the assumption that drivers follow the shortest path between their origin and their destination, but that one has to calculate an equilibrium where ``no driver could reduce his or her travel time by selecting a different route'' \cite{Wardrop1952}. This statement has been modified by shifting from travel time to ``generalized costs'' as decisive quantity for the equilibrium calculation, but nevertheless travel time in general has a heavy weight within the generalized costs. It is probably needless to say that the travel time is heavily influenced by the distribution of all the other participants of traffic, be it vehicular or pedestrian traffic.

There has been much less awareness for this issue on the side of pedestrians. This might be because pedestrians move not on a network as vehicles do, but freely in two spatial dimensions. As an effect micro simulation models of pedestrians are computationally more costly than simulations of vehicular dynamics and macro models of pedestrian dynamics are more difficult to be formulated. In result the iterative approach to finding an equilibrium (be it macro or micro) appears to be a tough case.

There are however situations in which travel time matters a lot for pedestrians, which is why they must base their movement decisions on the criterion which direction at some given point in time appears to promise the smallest remaining travel time. Pedestrians hustling through a station hall as they are late for a train have already been mentioned. Another situation would be an infrastructure providing two differently long paths through two distinct bottlenecks separated by at least a few meters. If the demand exceeds the capacity of the shorter path, but is below the capacity of both paths, after some time the jam on the shorter path will be large enough to produce delays that will or at least could make pedestrians familiar with the place detour on the longer path. As the geometry of the longer path can be arbitrarily complicated, this example makes it immediately clear that a model of pedestrian dynamics, which is based on movement along the shortest path plus elements that make pedestrians evade each other can never -- no matter what the basic approach and construction principle of the model is -- in general produce the desired behavior.

Following this line of motivation, we explicitly note here that the proposed method is not seen as a {\em general improvement} of the Social Force Model for all movement situations. It is rather an {\em alternative} way to calculate the direction of the desired velocity. Depending on the situation this can yield more realistic results, in others less realistic results, and in many cases similar results as a calculation based on the assumption of movement on the shortest path. In the latter case (comparable results) it is better to not apply it, as it is computationally more costly.

\subsection{The Social Force Model}
For a general overview on simulation of pedestrian dynamics and its history, as well as other modeling approaches apart from force based see \cite{Schadschneider2009}. In this subsection we focus on a short introduction of the Social Force Model itself.

The Social Force Model is one of the most discussed models of pedestrian dynamics. Since it was introduced originally by Helbing et al. \cite{Helbing1995,Helbing2000b} Helbing and members of his group have proposed a number of extensions \cite{Werner2003,Johansson2007,Helbing2009,Moussaid2010}, but also other authors have developed their own ideas based on the Social Force Model \cite{Yu2005,Lakoba2005,Pelechano2007,VISSIM2010,Steffen2010,Chraibi2010,Zainuddin2010,Kretz2011d}\footnote{Here the list of references inevitably has to be incomplete.}

It is a bit of a surprise that discussions of the Social Force Model more often deal with the forces between the agents than with the driving force term, although it is the driving force that sets the basic dynamics and although without the driving force one would be faced only with a diffusion process. Especially one particular issue has received only few attention in introducing as well as improving publications: {\em how is the direction of the desired velocity calculated?} With $v_{0}$ as desired speed, $\hat{v}_{0}$ as direction (unit vector) of the desired velocity, $\vec{v}$ as current velocity, and $\tau$ as inertia time parameter, the basic equation is
\begin{equation}
\vec{F}_{driving} = \frac{v_{0}\hat{v}_{0}-\vec{v}}{\tau}
\end{equation}

The reason for this is probably that in (geometric) models built to investigate the properties of the (dynamic) model, the desired direction usually is obvious. Typical geometries to test these properties -- like straight corridors -- do not need an elaborate method to calculate the direction of the desired velocity, as different methods of calculation would yield very similar directions. Studies in which this question has been addressed deal with panic situations and the question, if pedestrians follow others or their individual plan to reach the destination (on the shortest path) \cite{Helbing2000b,Lakoba2005,Zainuddin2010}, something which lies entirely outside the scope of this work\footnote{Actually these ``panic mechanisms'' trigger an opposite effect of what is intended in this paper: if all of a group of pedestrians desire to walk on the quickest path, the group typically spreads out and space is efficiently utilized. The panic modifications at the direction of the desired speed on the contrary typically make everyone head in a more similar direction than without these elements. This discrepancy is one source for the ``faster is slower'' and ``freezing by heating'' effects.}. There are exceptions from this focus on panic in dealing with the direction of the desired velocity as for example \cite{Freialdenhoven2010,Patil2010}, and very recently \cite{Moussaid2011}.

An obvious first idea for the direction of the desired velocity in arbitrary geometries is that it points into the direction of the shortest path to the destination. This can be achieved either by calculating a navigation graph -- probably the best of which is the visibility graph \cite{DeBerg1997} -- or by calculating a distance map (a.k.a. ``look up table of distances'', ``static potential'' or ``static floor field'') \cite{Burstedde2001} and receive the direction by calculating the gradient of the potential. We will make use of the distance map approach in the following and generalize the ``distance'' to an estimated remaining travel time.

The desired direction continuously can take any value between $0^\circ$ and $360^\circ$. The proposed numerical procedure implies some discreteness in the choices. However, the direction choice method proposed in this paper is still continuous in the sense that {\em no} geometric analysis is undertaken that creates a navigation graph on which discrete choices (e.g. ``pass an obstacle to the left or right?'') are made \cite{Bluemel2008,Hoecker2010,Guy2010}; no spatial semantics with rooms and doors or links and nodes is produced.

\subsection{Distance Maps}
A distance map can be calculated in a number of ways, often trading precision for computation time \cite{Kretz2010a}. Very fast na{\"i}ve flood fill methods can only result in metrics with norm $p=1$ or $p\rightarrow\infty$. 

\begin{equation}
||\vec{x}||_p = := \left(\sum_i |x_i|^p \right)^\frac{1}{p}
\end{equation}

If one wants a nearly exact method -- i.e. distance in Euclidean metric (vector norm $p=2$) under consideration of obstacles -- the distance map can be computed by solving the Eikonal Equation \cite{Bruns1895,Frank1927}
\begin{equation}
|\nabla S(\vec{x})|^2 = \frac{1}{f^2}.\label{eq:eikonalS}
\end{equation}
$S(\vec{x})$ is the distance map (shortest distance to the destination area under consideration of obstacles). For a distance calculation $f$ is a constant to receive the distance with the desired units. Here we are only concerned with the direction of the gradient of the resulting field, which does not depend on units or a global factor of the field and may choose $f\equiv1$ for walkable areas and $f\rightarrow 0$ for areas obstructed by obstacles. 

Equation (\ref{eq:eikonalS}) alone is not sufficient but one needs to add a boundary condition which fixes the values of $S$ on the edge of the destination area (calling the edge of the destination area $\delta A_d$, it is usually defined $S(\delta A_d) = 0$). Inside obstacles $S$ is initialized with $\infty$ -- implemented in a computer program by using a numerical value with equivalent effect. 

``Solving the Eikonal Equation numerically'' can be imagined as placing a discrete lattice in the background, begin at the grid points of $\delta A_d$ and drive outward a front which sets the values of $S$ on all other grid points by summing up the distances between the grid points over which it is ``flowing''. Simple flood fill methods result in Manhattan metric (vector norm $p=1$) or Chessboard metric (vector norm $p\rightarrow\infty$). Desired and to be called ``error free'' is the Euclidean metric (vector norm $p=2$). Numerical methods to solve the Eikonal Equation efficiently, which result in a Euclidean metric, are for example the Fast Marching Method \cite{_Sethian1999} and the Fast Iterative Method\cite{Jeong2007}.

In our simulations we use a lattice point spacing between 15 and 20 cm, i.e. the lower range of human body diameters. Larger lattice point spacings lead to unrealistic artifacts and produce problems at narrow building infrastructure, smaller lattice point spacings do not yield advantages, but occupy more comßputer memory. 

No matter if a na{\"i}ve flood fill method or an Eikonal Equation Solver is used, once $S$ has been calculated, the direction of the desired velocity is obtained from the gradient of $S(\vec{x})$: 
\begin{equation}
\vec{v}_0 = -v_0\frac{\nabla S(\vec{x})}{|\nabla S(\vec{x})|}
\end{equation}
where the desired speed $v_0$ is an external parameter to be set by the scenario modeler and technically ``gradient'' is to be understood as discrete gradient, as the field $S$ is not continuous but only defined on the points of a grid.

With the motivation of this paper, it is clear that a ``distance map'' cannot be the end of the story. What we need is a ``map of estimated remaining travel times'' or at least a map which is different from such a map just by a global factor. Mathematically it is only a small step from the computation of the shortest to the quickest path: in equation (\ref{eq:eikonalS}) $f$ is not to be taken constant, but proportional to the speed expected at that spot. If the travel speed is expected to be the same everywhere then the estimated remaining travel time is different from the distance to destination only by a global factor (which is the speed). However, if at different spots different travel speeds are expected, the value of $f$ at the spots needs to take this into account: $f$ is no longer a global constant, but it is a grid (a field) itself, a grid of expected travel speeds.

Recently it has been shown that the field of gradients of the distance map can be calculated without explicitly calculating the distance map \cite{Schultz2010a}. The method might even be faster than the fastest Eikonal Equation Solver, yet it is not suited to compute the gradient field of a map of travel times.

Normally this would be the place to have a discussion of preceding work, but as with some of these an in-depth discussion of the details is done, it appears that it makes more sense if first the model extension is introduced.

\section{Calculating the Direction of the Quickest Path}\label{sec:Method}

\begin{figure}[htbp]
\begin{center}
	\includegraphics[width=\figurewidth]{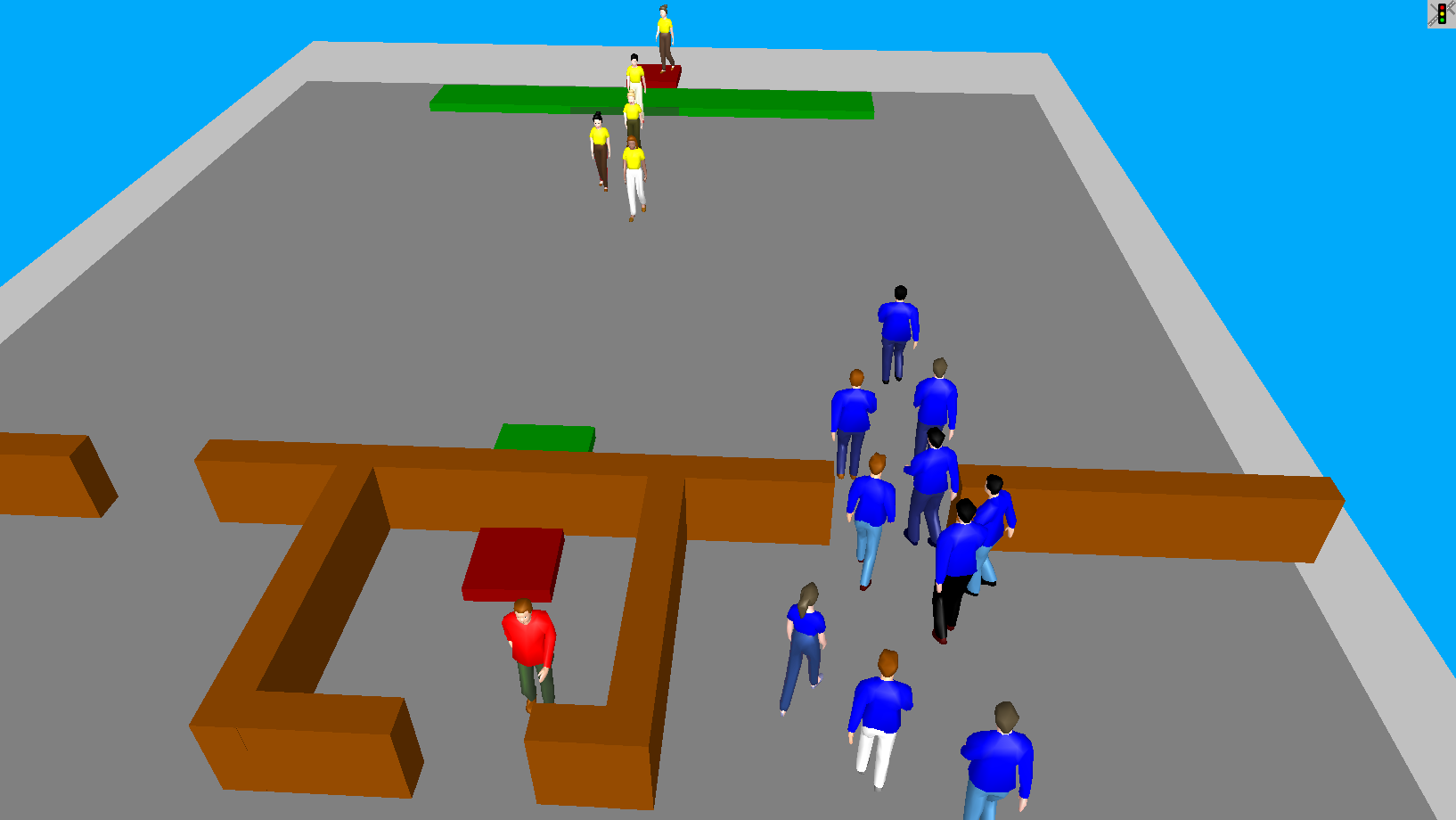} \\ \vspace {12pt}
	\includegraphics[width=\figurewidth]{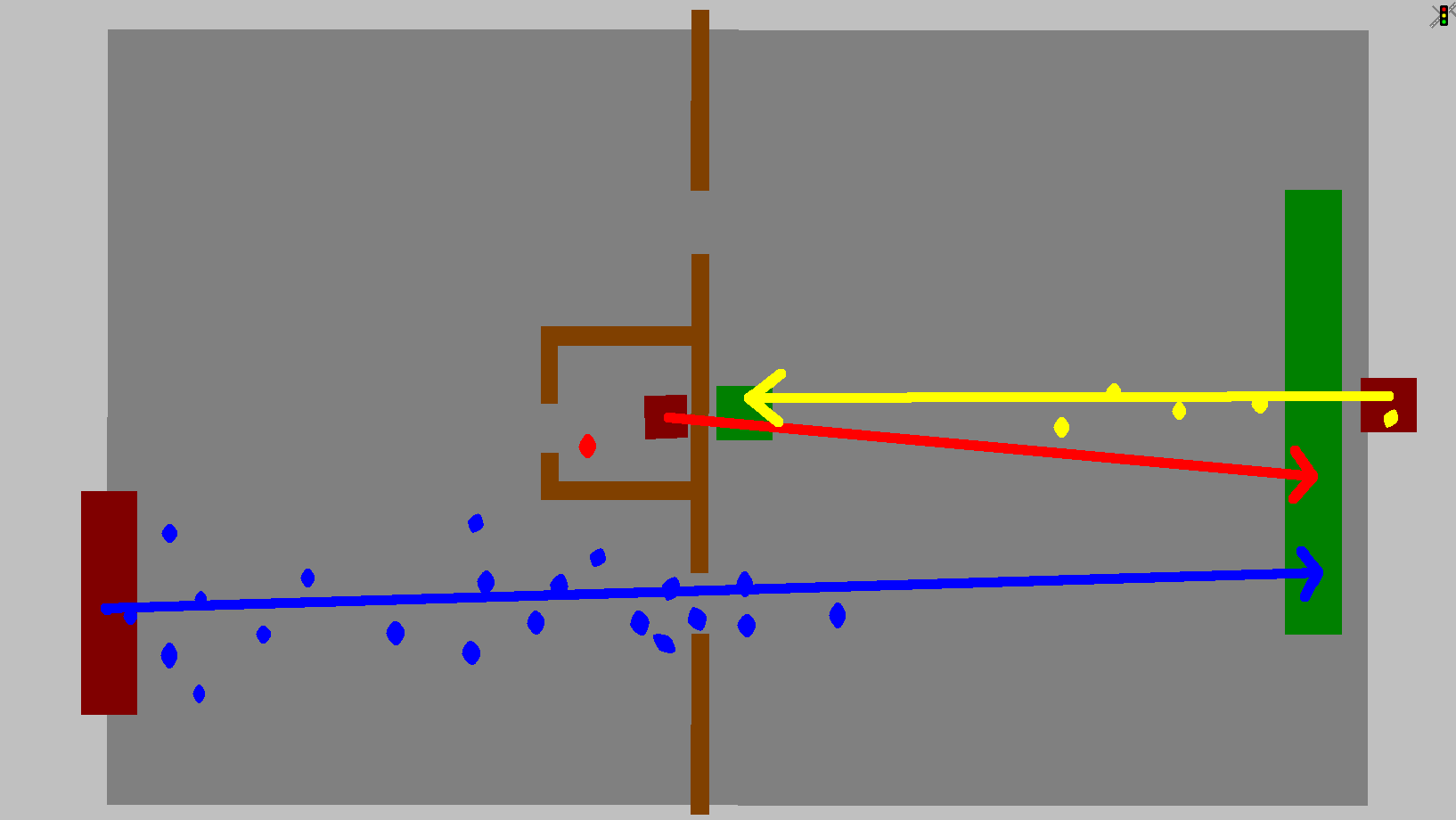}
\caption{{\bf 3D view (upper figure):} Example scenario to demonstrate the method and its effect. The blue and red agents are heading for the remote green area. This is the destination area for which the dynamic potential is calculated. There are two openings in the wall that separates the two rooms. It is not defined which opening the red agents are to use. The center of the exit of the room where the red agents have their origin is about 80 cm closer to the door to the right (which is used by the blue agents) than to the door to the left. The number of red agents being inserted into the simulation is much smaller than the number of blue agents. The yellow agents are just added to demonstrate the effect of opposing agents on the dynamic potential and are walking not through one of the doors of the wall, but just to the small green area. {\bf 2D view (lower figure):} Additionally origin (dark red) and destination (dark green) for each color of agents are linked with an arrow of corresponding color. Agents are set into the simulation somewhere on the starting area randomly and with equal probability; it is taken care, though, that, no collision at input occurs. The edge of the green destination areas in all cases and at all times is of equal value for the corresponding agents of a simulation run. Figures \ref{fig:screen4oneoverf} and \ref{fig:oneoverf} show the agents' impact on the field of $f$.}
\label{fig:3d-overview} 
\end{center}
\end{figure}

To calculate the current direction of estimated least travel time (shorter: ``dynamic potential'') comprises of three steps:

First a map of expected or estimated walking speeds for small areas must be calculated which takes account for the distribution of obstacles, agents and other properties like walking surface quality that may influence speed. If -- as in this work -- the small areas are tiles of a regular grid, the inverse of the estimated walking speeds is proportional to the travel times over the tiles. This is the traffic science part of the method.

Second: Beginning at the destination area all the travel times of the tiles are numerically integrated using a numerical Eikonal equation solver. The result is the desired field of estimated remaining travel times to the destination area. For each grid point there is now such a value available for further usage in the pedestrian dynamics model. This is the mathematical part of the method. It also presents the major challenge for the implementation of the method, as for most models of pedestrian dynamics this new method will imply a relevant additional amount of computation time. It is therefore advisable to think well about an efficient implementation.

The third step is to calculate the gradients in the dynamic potential for at least all grid points which are occupied by an agent which is heading for the destination of that particular dynamic potential.

\subsection{Step One: Estimating Walking Speeds}

As noted above, this contribution will only deal with the effect of the distribution of the agents on the estimated walking speed. As the relevant result -- the field of gradients -- is invariant to a global factor on the dynamic potential, for any unoccupied grid point at walkable space the value of $f$ for that grid point is set unit-less to $f=1$. If on the contrary a grid point at walkable space is occupied by an agent, $f$ is set to a value $f\leq1$ according to:

\begin{equation}
\frac{1}{f} = 1+\max \left(0,g\left(1+h\frac{\vec{v}\cdot\nabla S}{v_0|\nabla S|}\right)\right)\label{eq:f1}
\end{equation}

$S$ is the distance map of the corresponding destination, the gradient is taken at the corresponding grid point. $\vec{v}$ is the current velocity of the agent that occupies the grid point and $g$ and $h$ are free parameters of the method: $g$ sets the general impact strength of the method ($g=0$ means $1/f=1$, i.e. the dynamic potential becomes a static map of distances) and $h$ sets the impact of the moving direction of an agent. For further considerations on the role and numerical values of $g$ and $h$ see subsection \ref{subsec:gh}. 

It would be desirable to calculate a dynamic potential for each agent individually. Then $v_0$ would be the desired speed of the agent for which the dynamic potential is calculated. However, for simulation scenarios of reasonable size this would imply unacceptable computation times and memory demand on computers as they are available off-the-shell today. Until this has changed the simplest compromise is to calculate one dynamic potential for all agents heading to the same destination and use as value for $v_0$ the average of desired speeds of all these agents. The implications of this compromise and possible more elaborate compromises are discussed in subsection \ref{subsec:OpenIssues}.

At this point there might be some confusion: one might think that using the Eikonal Equation to calculate temporal distances can easiest be achieved, $f$ in the right hand side of equation  (\ref{eq:eikonalS}) is set to be the current speed of a pedestrian for all locations which are occupied by pedestrians heading for this destination. However, this simple approach cannot work out, as the example of an agent at rest shows: this would result in $1/f=1/0$. In principle that could be handled in the algorithm by handling agents at rest exactly as if they were solid, static obstacles.  Then, however, it could happen that agents at rest block the numerical integration before all grid points have been assigned a value for the estimated remaining travel time. A line of agents at rest from one wall of a corridor to the opposing one would result in an infinitely large travel time through this corridor, which is an unrealistic estimation. Therefore the speed of the agent at the grid point cannot be used directly, but an equation as (\ref{eq:f1}) is needed, which prevents $1/f$ from diverging. A second reason against directly using the current speed of the agent as value for $f$ is that the agent occupying the spot might head for an entirely different direction, even opposing all the agents that are longing for the destination for which the dynamic potential is calculated. Then, even if their speed is desirable, they are counted to probably cause a specially large delay for agents that might have to pass that spot on their way to the destination of the dynamic potential. Finally a third reason why a measured speed cannot be used as value for $f$ -- even in uni-directional movement to one single destination -- is the experience in macroscopic traffic planning: high demand implies jams, jams imply vehicles which are at rest at certain times, nevertheless do the common capacity restraint functions not have a pole, i.e. travel time grows fast but infinitely large only with infinitely large demand (i.e. there is no singularity in the function) \cite{helbing2009b}, which for the matter of space requirement of a single vehicle is not possible.

\begin{figure}[htbp]
\begin{center}
	\includegraphics[width=\figurewidth]{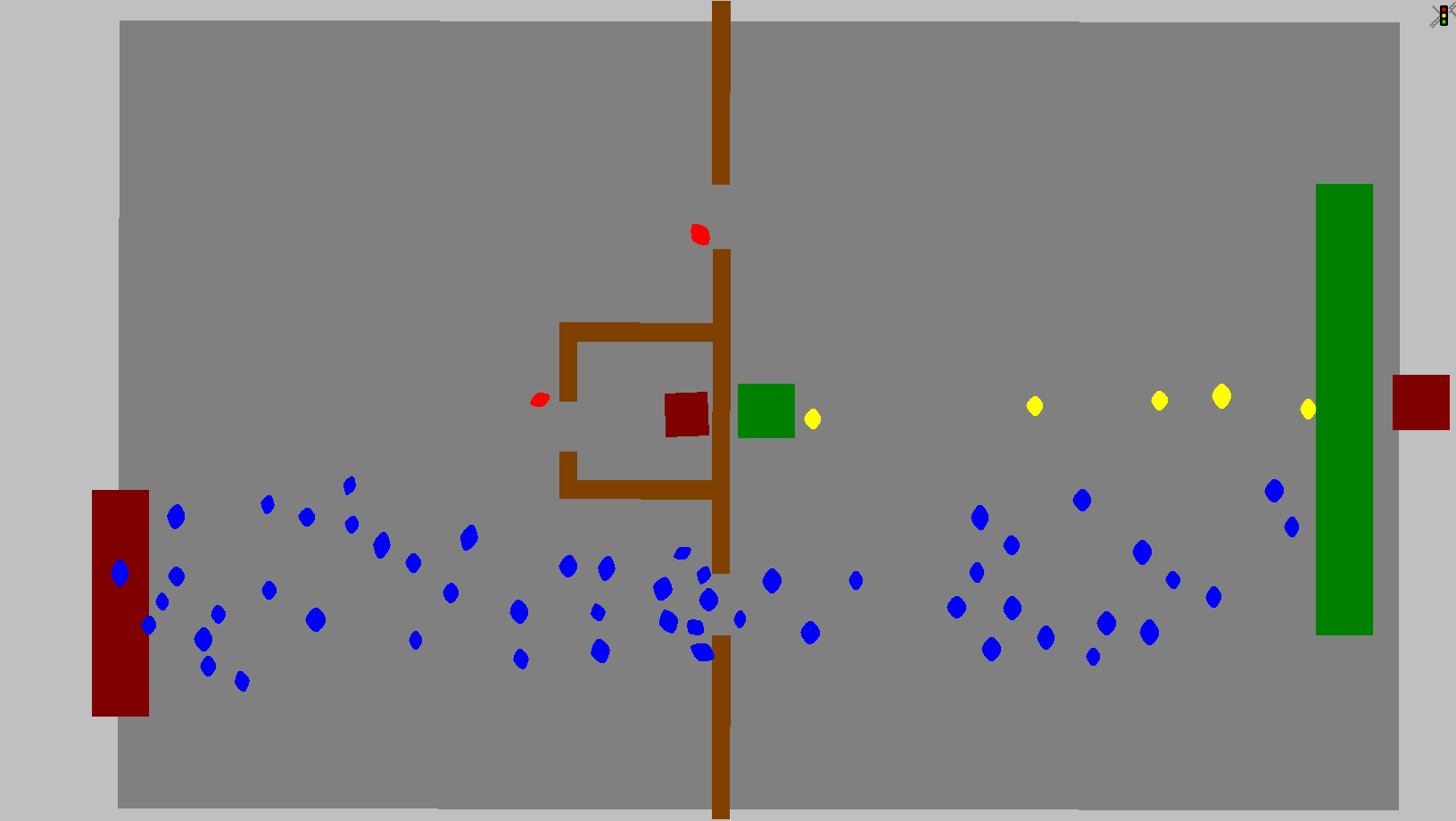}
\caption{A situation during the simulation. Figure \ref{fig:oneoverf} shows what this implies for the field of $1/f$.}
\label{fig:screen4oneoverf} 
\end{center}
\end{figure}

To develop an intuition of equation (\ref{eq:f1}) assume that agent $j$ is influenced by a dynamic potential and $j$ is somehow behind (upstream) another agent $i$ and both are heading into the same direction toward the destination along the shortest path, i.e. $\vec{v}_i$ and $\nabla S$ point into opposite directions (the gradient always points upstream, away from the destination). Further set $h=1$ for this example. Then equation (\ref{eq:f1}) simplifies to

\begin{equation}
\frac{1}{f} = 1+\max \left(0,g\left(1-\frac{|\vec{v}_i|}{v_0}\right)\right)\label{eq:f2}
\end{equation}
for the value for $1/f$ at the position of agent $i$.

If $i$ is faster than $j$ wants to walk ($v_i>v_0$; $v_0$ is the desired speed of the agent who is influenced by the dynamic potential, i.e. agent $j$) then there is no need for $j$ to deviate from the shortest path and consequently equation (\ref{eq:f2}) gives $f=1$, as if the spot, where agent $i$ is located was unoccupied. But if $j$ wants to walk faster than $i$, at some point $j$ needs to start an overtaking maneuver, i.e. evade from the shortest path and indeed in this case (\ref{eq:f2}) gives $f<1$. If agent $i$ would even oppose agent $j$ and walk exactly upstream in the static potential $S$ heading for some different destination, then (\ref{eq:f1}) would read

\begin{equation}
\frac{1}{f} = 1+g\left(1+\frac{|\vec{v}_i|}{v_0}\right). \label{eq:f3}
\end{equation}
for the value for $1/f$ at the position of agent $i$.

For a speed of $v_i$ this is the largest value $1/f$ can take, i.e. agent $i$ with its presence and movement exerts the maximum possible effect on the value of $f$.

\begin{figure}[htbp]
\begin{center}
	\includegraphics[width=\figurewidth]{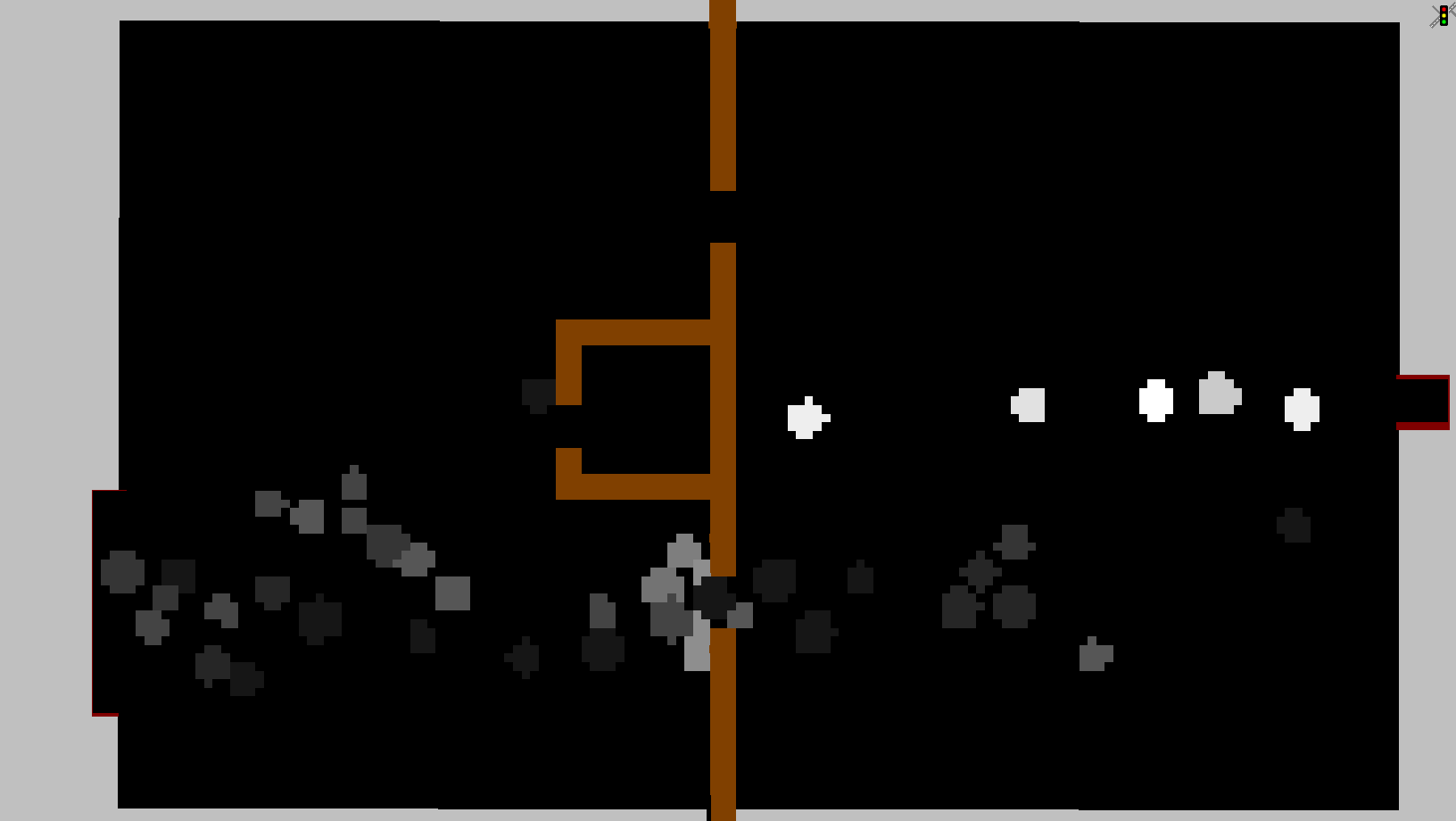}
\caption{This figure shows the field $1/f$ for the agents (blue and red ones) heading to the right destination area (large green area in figure \ref{fig:screen4oneoverf}). The gray value of a spot reflects the impact of the agents as shown in figure \ref{fig:screen4oneoverf} on the field of $f$. In entirely black areas we have $f=1$ and the gray value scales with $1/f$, i.e. the brighter a spot is, the larger is the expected travel time delay: it can clearly be seen that opposing agents have the strongest impact (brightest spots). This figure shows that the area affected by an agent exceeds the area which is actually occupied. The radius of the affected area is increased by 50\% compared to the radius of an agent to achieve approximately that the {\em center} of another agent is on an affected area as soon as the two agents would touch each other. The resulting dynamic potential is shown in figure \ref{fig:statpot}.}
\label{fig:oneoverf} 
\end{center}
\end{figure}

Figure \ref{fig:oneoverf} illustrates this velocity dependent impact of the agents on $1/f$.

As stated above for reasons of computational effort we only calculate one dynamic potential per destination. Therefore $v_0$ is the average desired speed of all agents being affected by that particular dynamic potential. This is discussed further below.

\subsection{Step Two: Numerical Integration for the Travel Time Map}

\begin{figure}[htbp]
\begin{center}
	\includegraphics[width=\figurewidth]{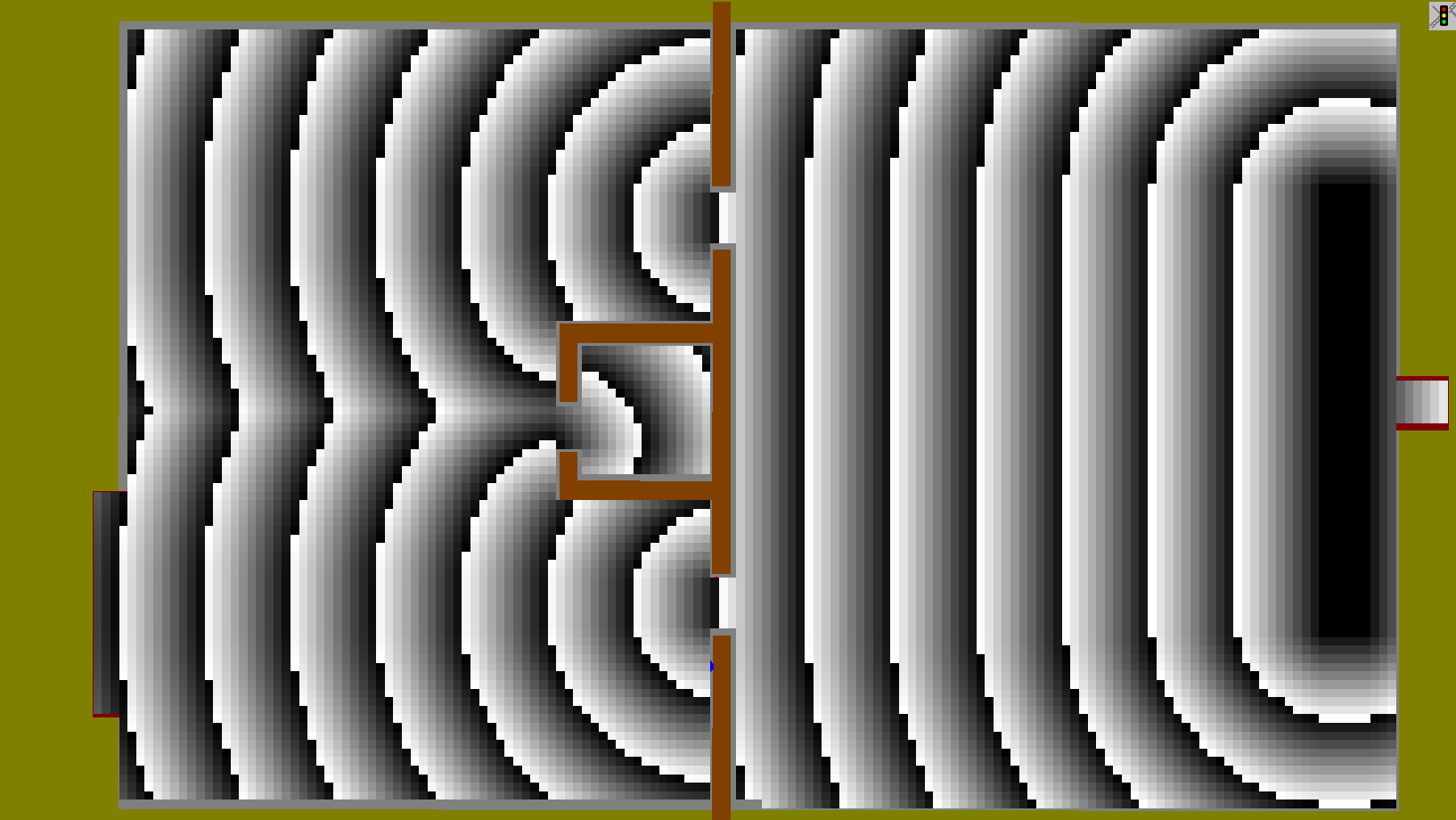} \\ \vspace{12pt}
	\includegraphics[width=\figurewidth]{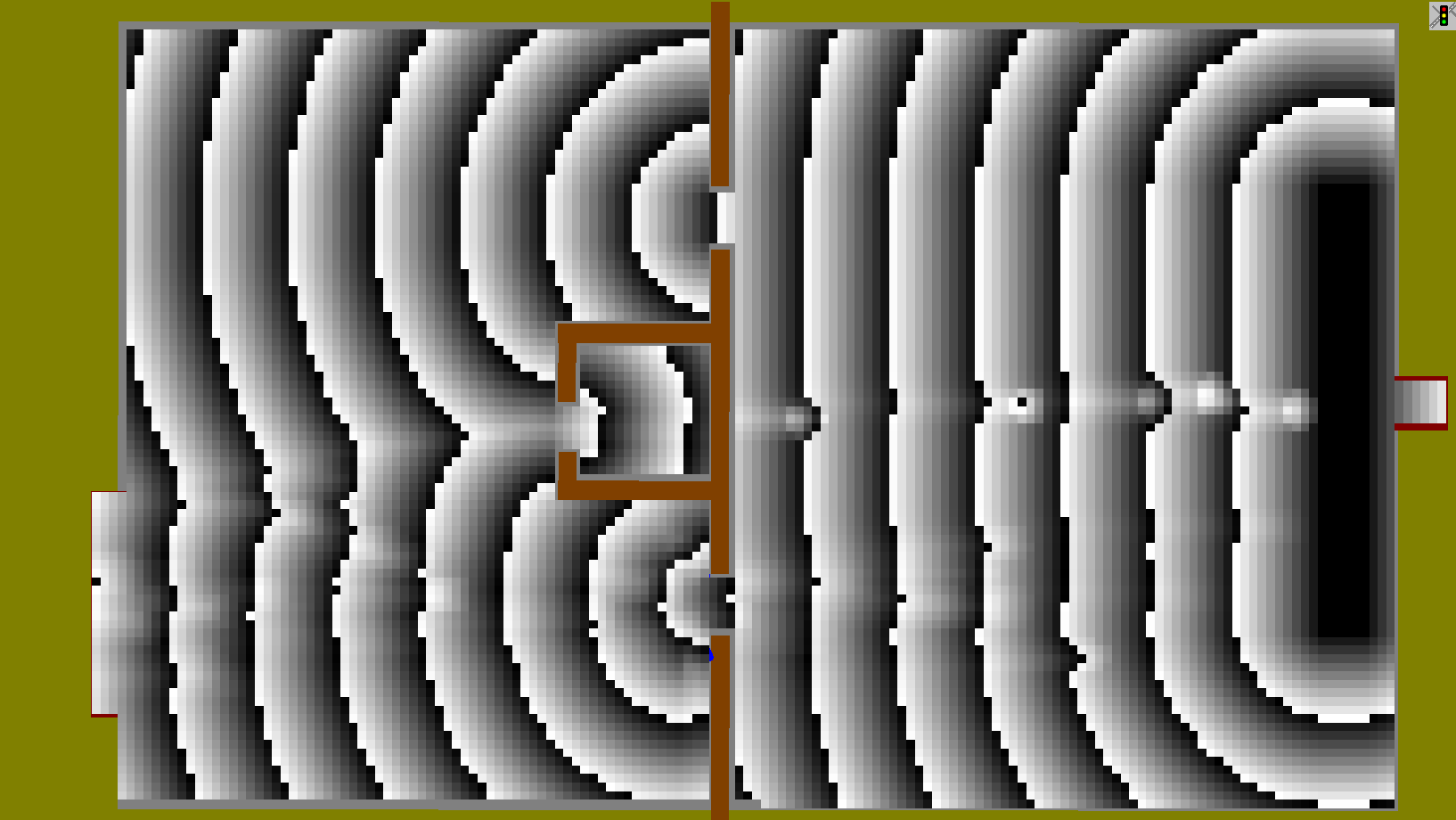}
\caption{Comparison of static potential / map of distances to destination (upper figure) and dynamic potential / map of estimated remaining travel times (lower figure) for the example scenario as shown in figure \ref{fig:screen4oneoverf}. In principle in this figure a brighter spot means that the value there is larger, but for better visibility the gray-scale is modulo{\"e}d. Lines of constant gray-value are lines of constant potential value. The gradients are always oriented orthogonal to these lines of constant value. The dynamic potential results when the field of $f$ as shown in figure \ref{fig:oneoverf} is numerically integrated. Note the impact of individual agents on the dynamic potential and how the dynamic potential appears to be ``compressed'' at the opening that is used by the blue agents. Figure \ref{fig:dyngrad} shows the field of gradients following from the dynamic potential.}
\label{fig:statpot} 
\end{center}
\end{figure}

In mathematical terms one receives the map $T$ of estimated travel times to the destination by solving the Eikonal Equation with the estimated speed $f(\vec{x})$ of an agent on a spot $\vec{x}$ on the right side:

\begin{equation}
|\nabla T(\vec{x})|^2 = \frac{1}{f(\vec{x})^2},\label{eq:eikonalT}
\end{equation}

Numerical methods to solve the Eikonal Equation efficiently and resulting in a Euclidean metric are the well-established Fast Marching Method (FMM) \cite{Kimmel1998,_Sethian1999}, which is widely used for various applications, and the less known and more recent Fast Iterative Method (FIM) \cite{Jeong2007,Jeong2007b,Jeong2008,Jeong2008b}. The major difference between both is, that the FMM has a better worst case computation time behavior, while the FIM algorithm is much more easy calculated in a multi-threaded way and therefore often has smaller computation times.

\begin{figure}[htbp]
\begin{center}
	\includegraphics[width=\figurewidth]{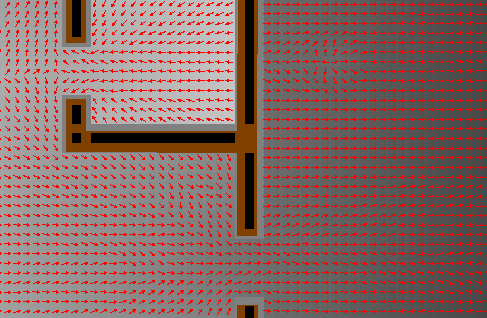}
\caption{A detailed view of the field of (negative) gradients as it follows from the dynamic potential of figure \ref{fig:statpot}.}
\label{fig:dyngrad} 
\end{center}
\end{figure}

Once the dynamic potential $T$ has been calculated, the direction of the desired velocity follows from it as the gradient:
\begin{equation}
\hat{v}_0=-\frac{\nabla T}{|\nabla T|}
\end{equation}

\begin{figure}[htbp]
\begin{center}
	\includegraphics[width=0.45 \textwidth]{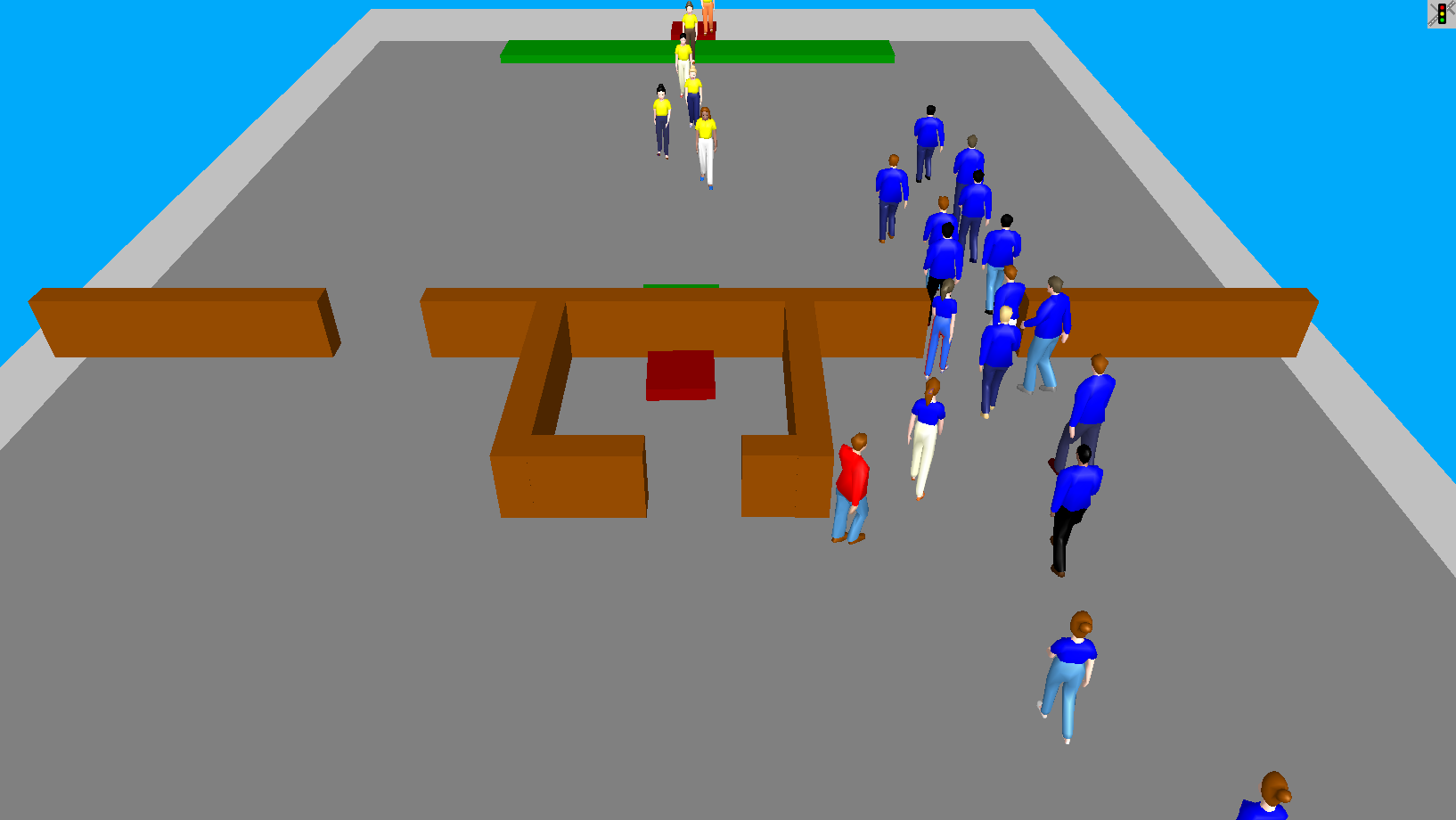}  \hspace{12pt}
	\includegraphics[width=0.45 \textwidth]{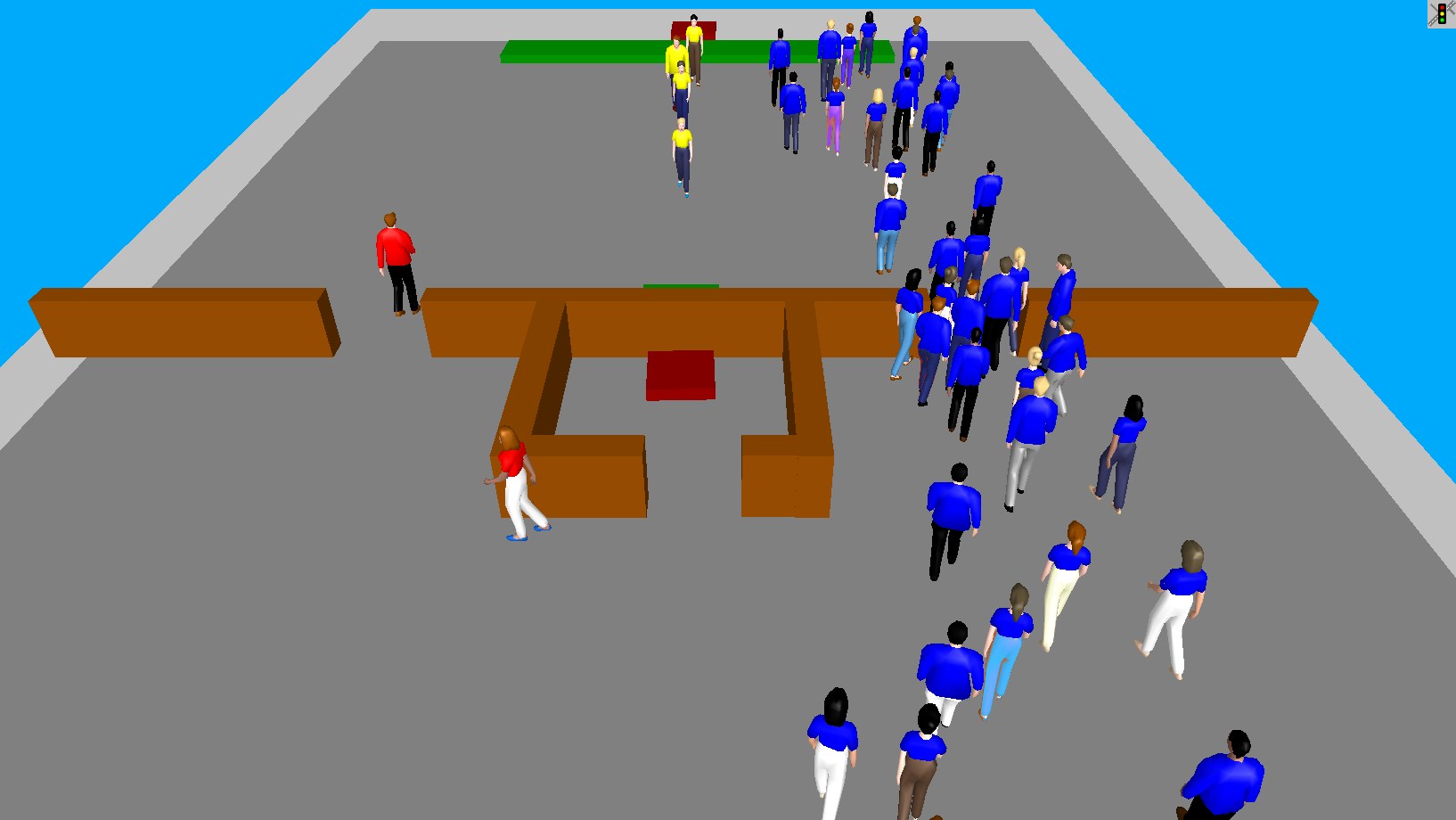}

\caption{Resulting behavior: while in the figure on the left the red agents walk to the (closer) right opening if there are only few blue agents, in the figure on the right side they prefer the left opening, if there is a jam in front of the right opening.}
\label{fig:right} 
\end{center}
\end{figure}

The mathematical process sketched in this subsection is illustrated in figures \ref{fig:statpot} to \ref{fig:right}.

\section{Examples} \label{sec:Examples}
\subsection{A U-Turn}
\begin{figure}[htbp]
\begin{center}
\includegraphics[width=0.45\textwidth]{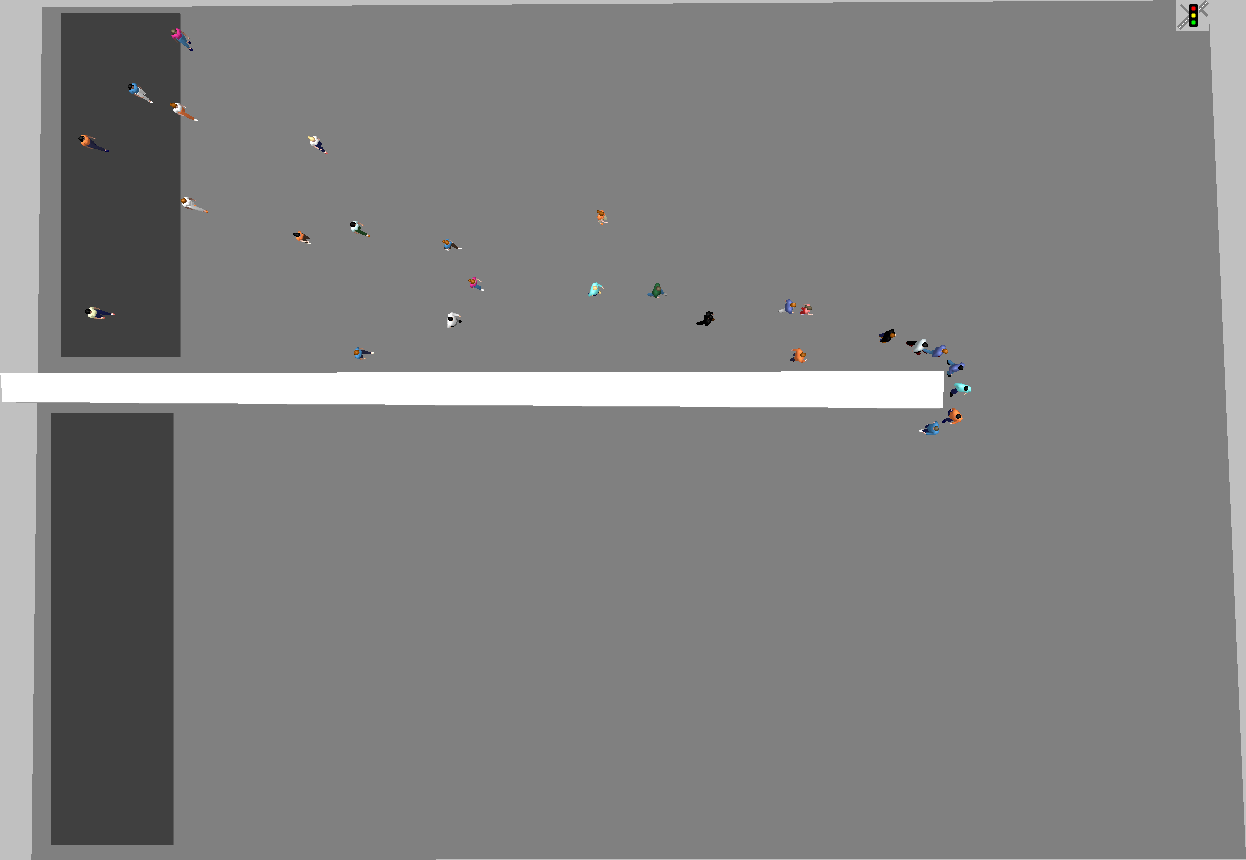} \hspace{12pt}
\includegraphics[width=0.45\textwidth]{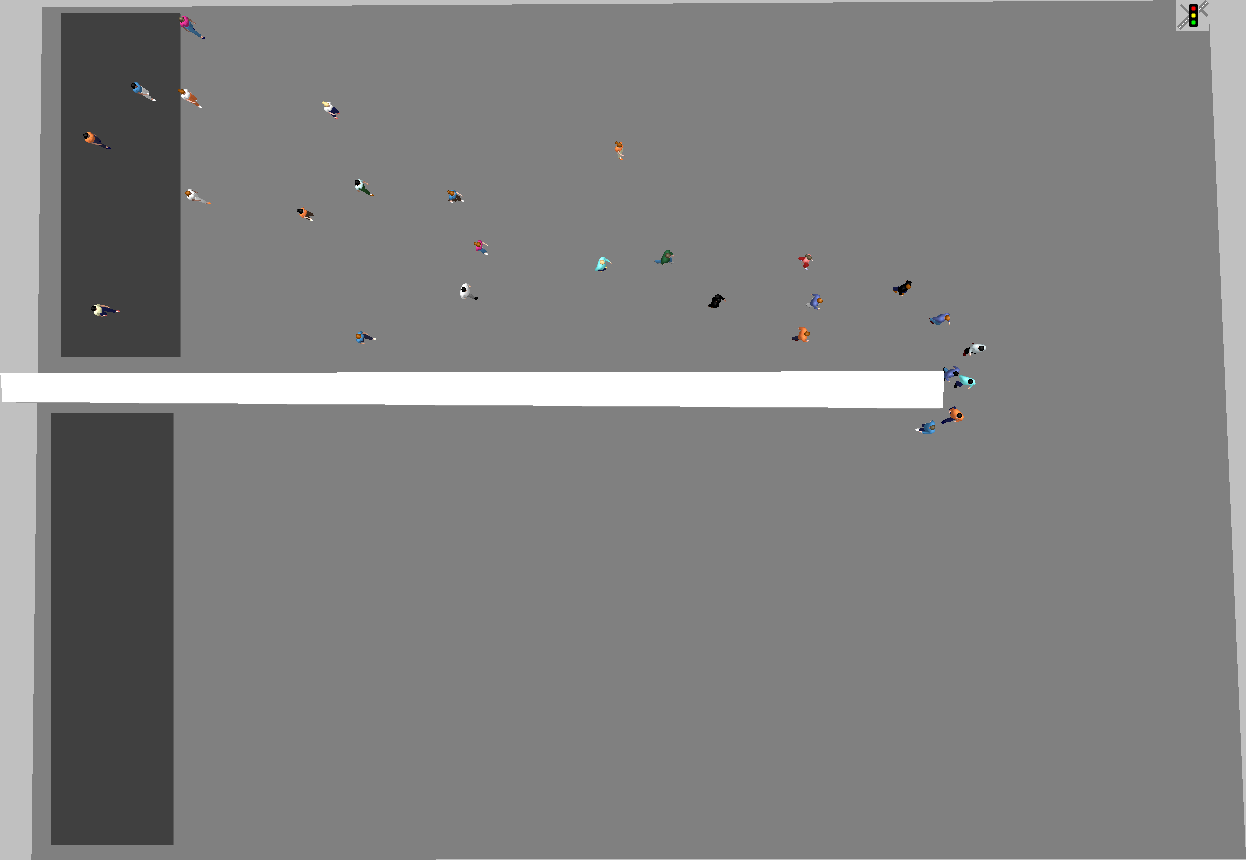}\\ \vspace{6pt}
\includegraphics[width=0.45\textwidth]{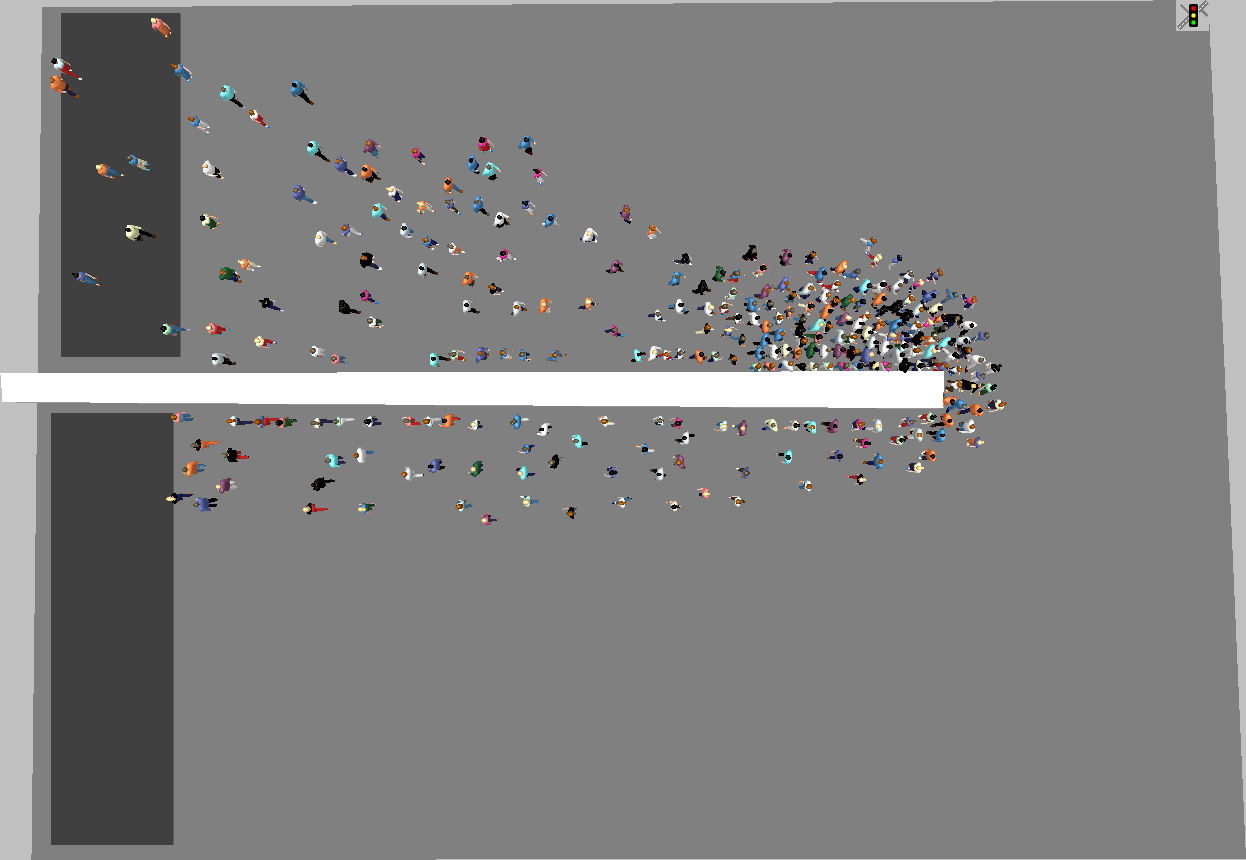} \hspace{12pt}
\includegraphics[width=0.45\textwidth]{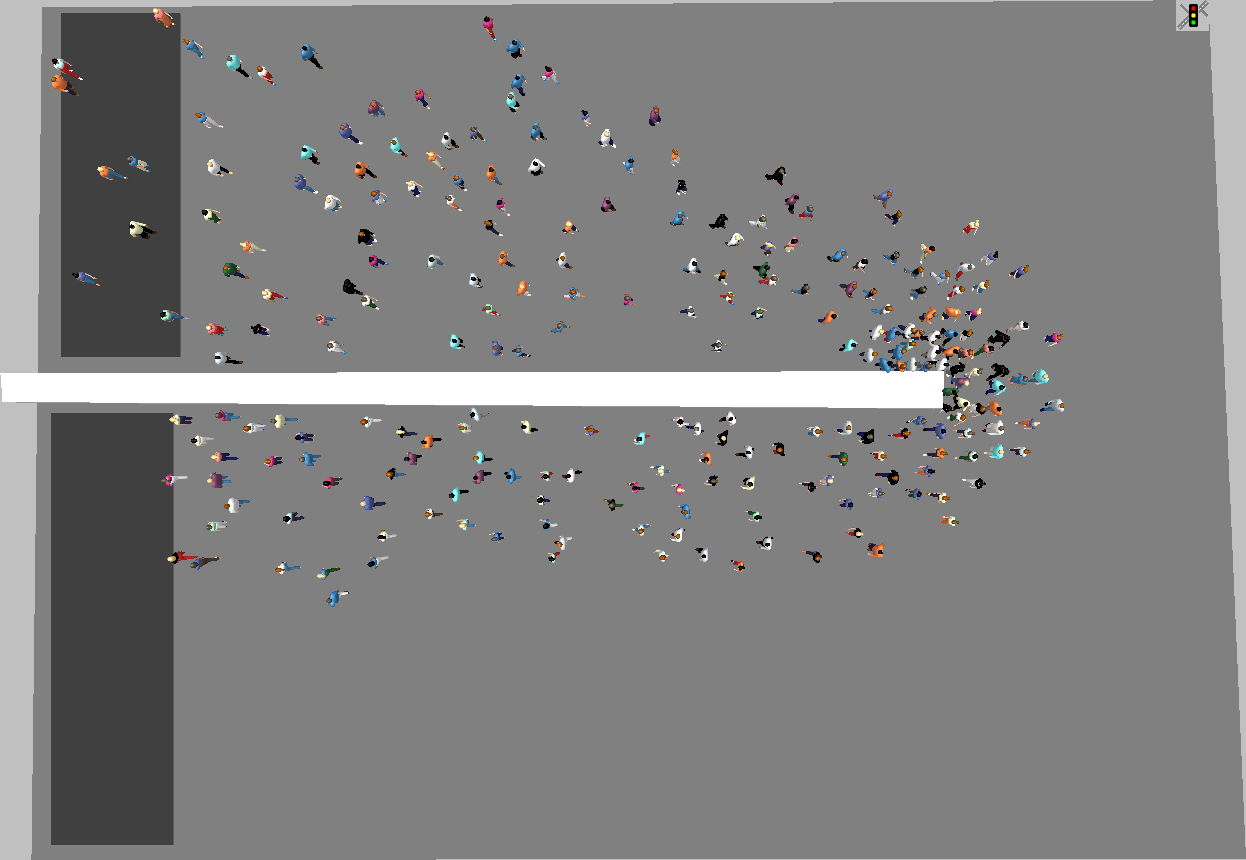}\\ \vspace{6pt}
\includegraphics[width=0.45\textwidth]{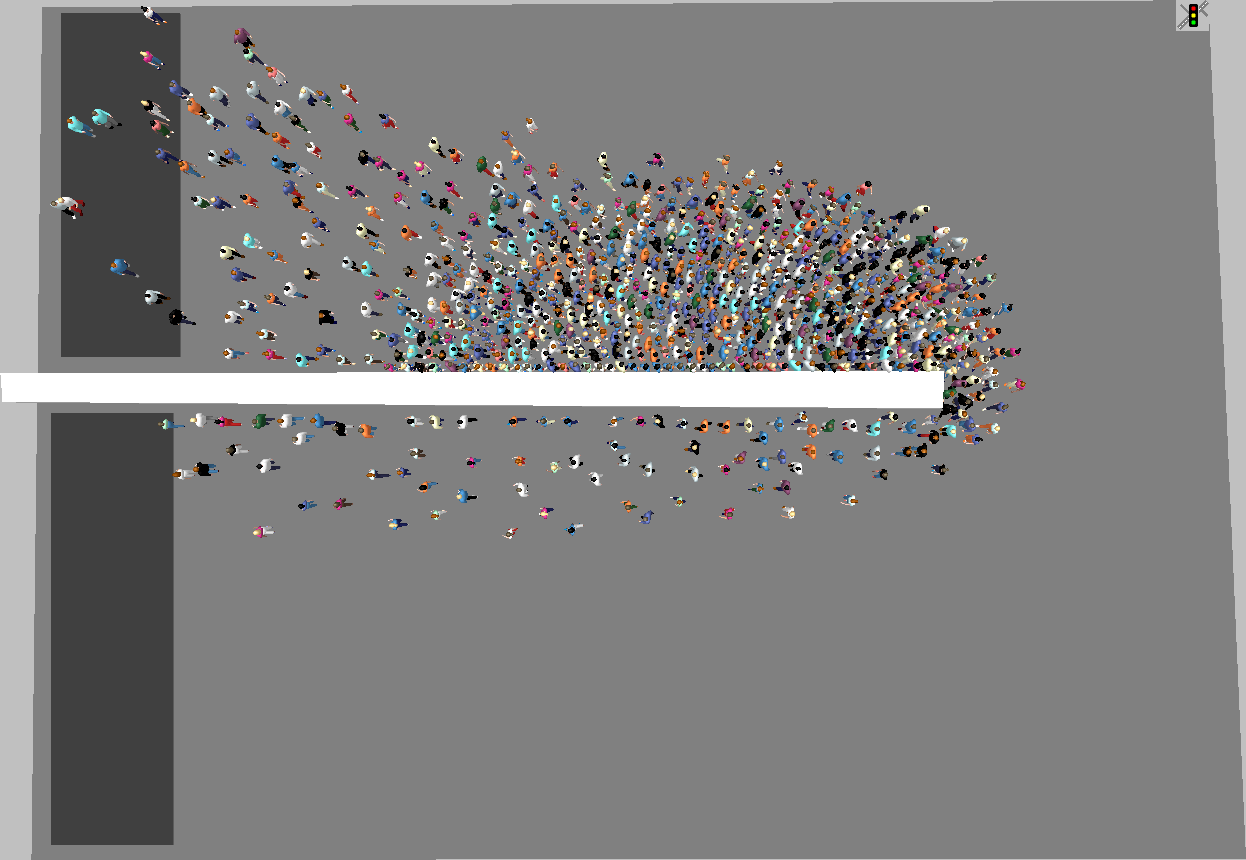} \hspace{12pt}
\includegraphics[width=0.45\textwidth]{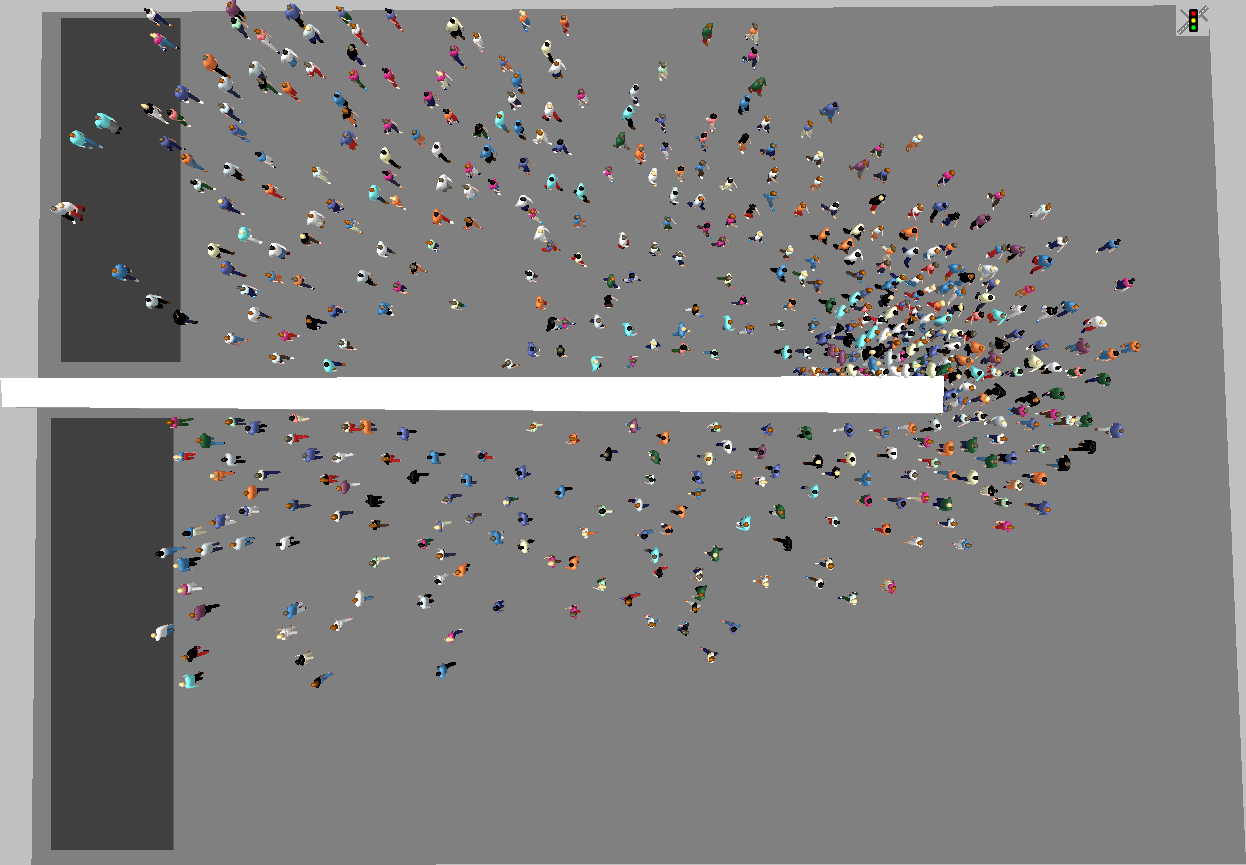}\\ \vspace{6pt}
\includegraphics[width=0.45\textwidth]{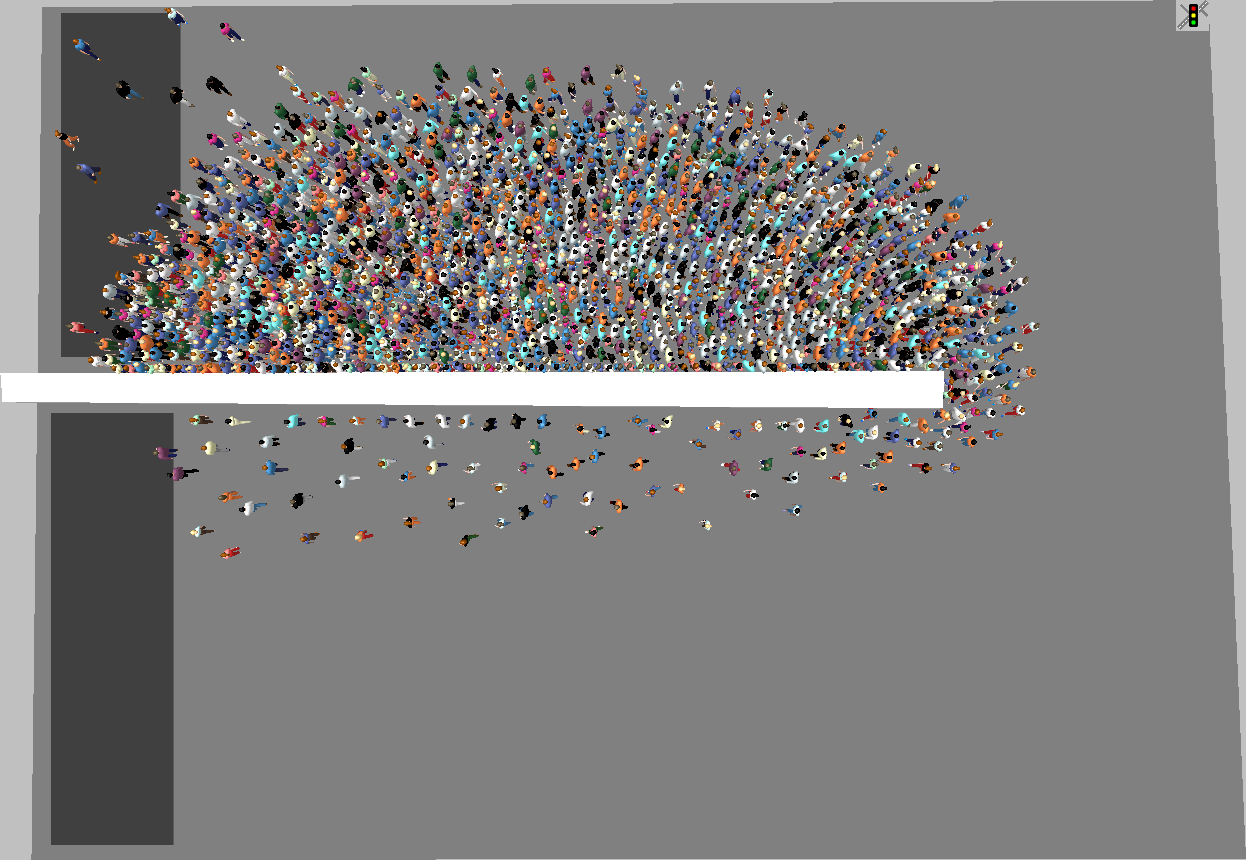} \hspace{12pt}
\includegraphics[width=0.45\textwidth]{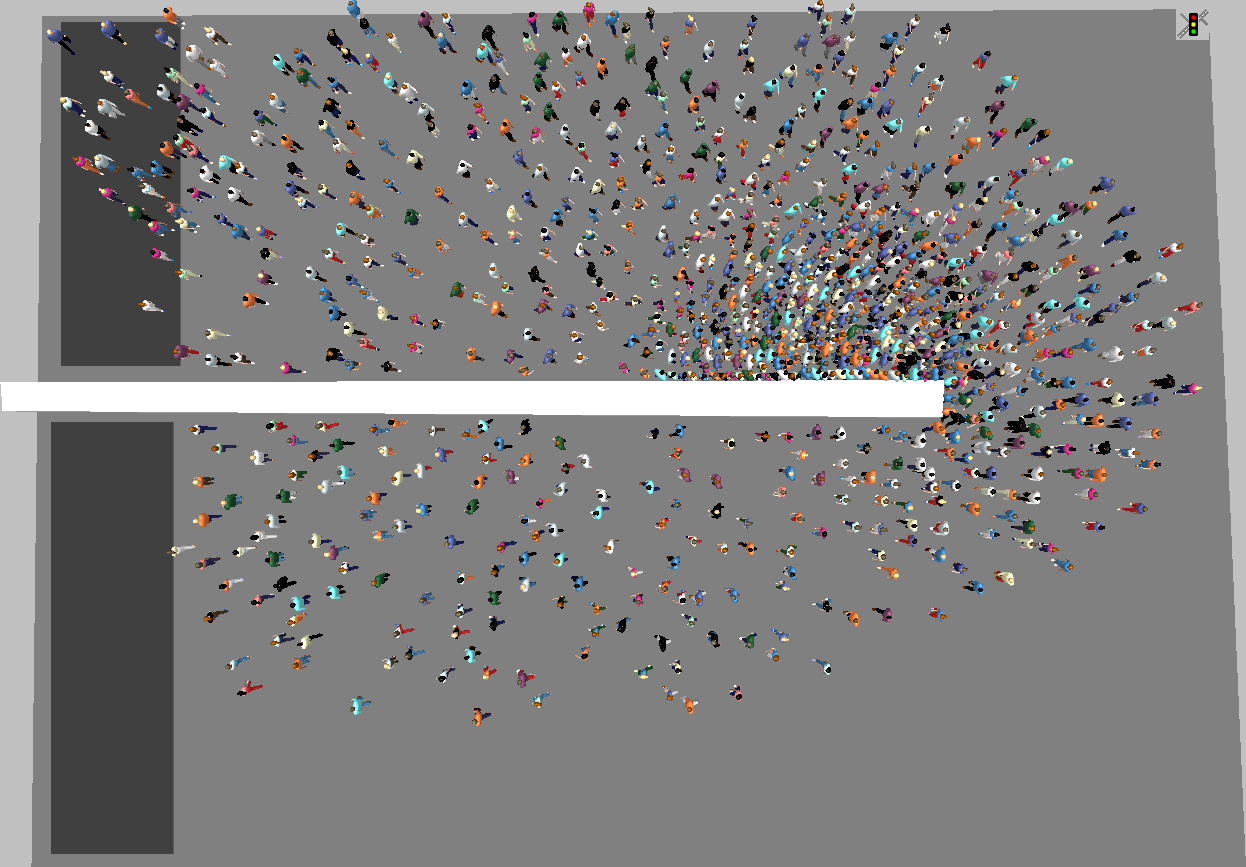}\\
\caption{Comparison of a crowd of agents walking around a 180 degree corner without (left column) and with (right column) dynamic potential. For this scenario $g=2.5$ has been set (a relatively large value). Scaling down to $g=0.0$ the behavior of the agents can continuously be transformed to the one of the static potential. The static and dynamic potential for the last two screenshots are shown in figure \ref{fig:U-Turn-Pot}.}
\label{fig:U-Turn}
\end{center}
\end{figure}

Figure \ref{fig:U-Turn} shows a crowd of agents walking around a u-turn. The demand increases over time from 1 to 7 agents per second.

\begin{figure}[htbp]
\begin{center}
\includegraphics[width=\figurewidth]{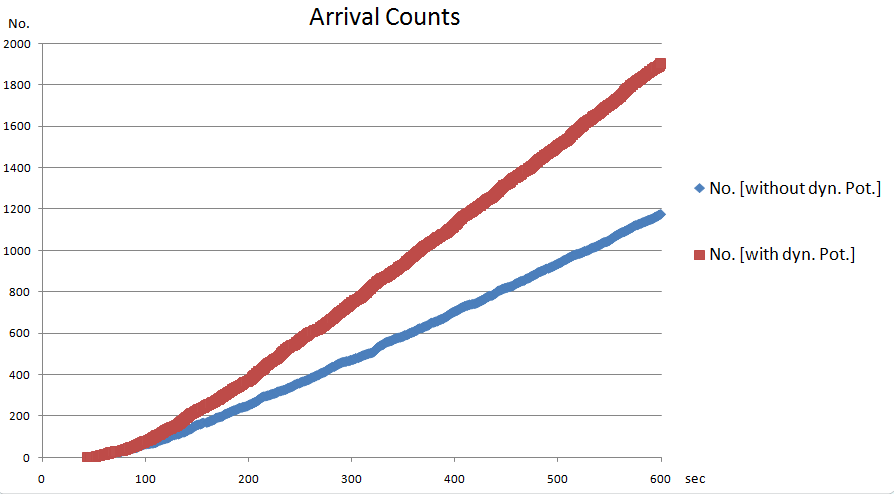} 
\caption{Number of agents that have a arrived at a given time. The input is on average 4 pedestrians per second, without dynamic potential the average arrival flow is about 2.35 agents per second, with dynamic potential it is 3.85 agents per second.}
\label{fig:U-Turn-Counts}
\end{center}
\end{figure}

The u-turn example shows why a sequence of destination lines as it has been investigated in \cite{Freialdenhoven2010} along the radii of the turn does not help to achieve the desired agent behavior: a sequence of destination lines would also make agents walk on larger radii around the corner when there is only a very low agent demand and density. If, however, only for example each ten seconds one agents walks around the corner, it should walk closely to the globally shortest path. Figure \ref{fig:U-Turn-Counts} compares the arrival flows without and with dynamic potential.

\begin{figure}[htbp]
\begin{center}
\includegraphics[width=0.45\textwidth]{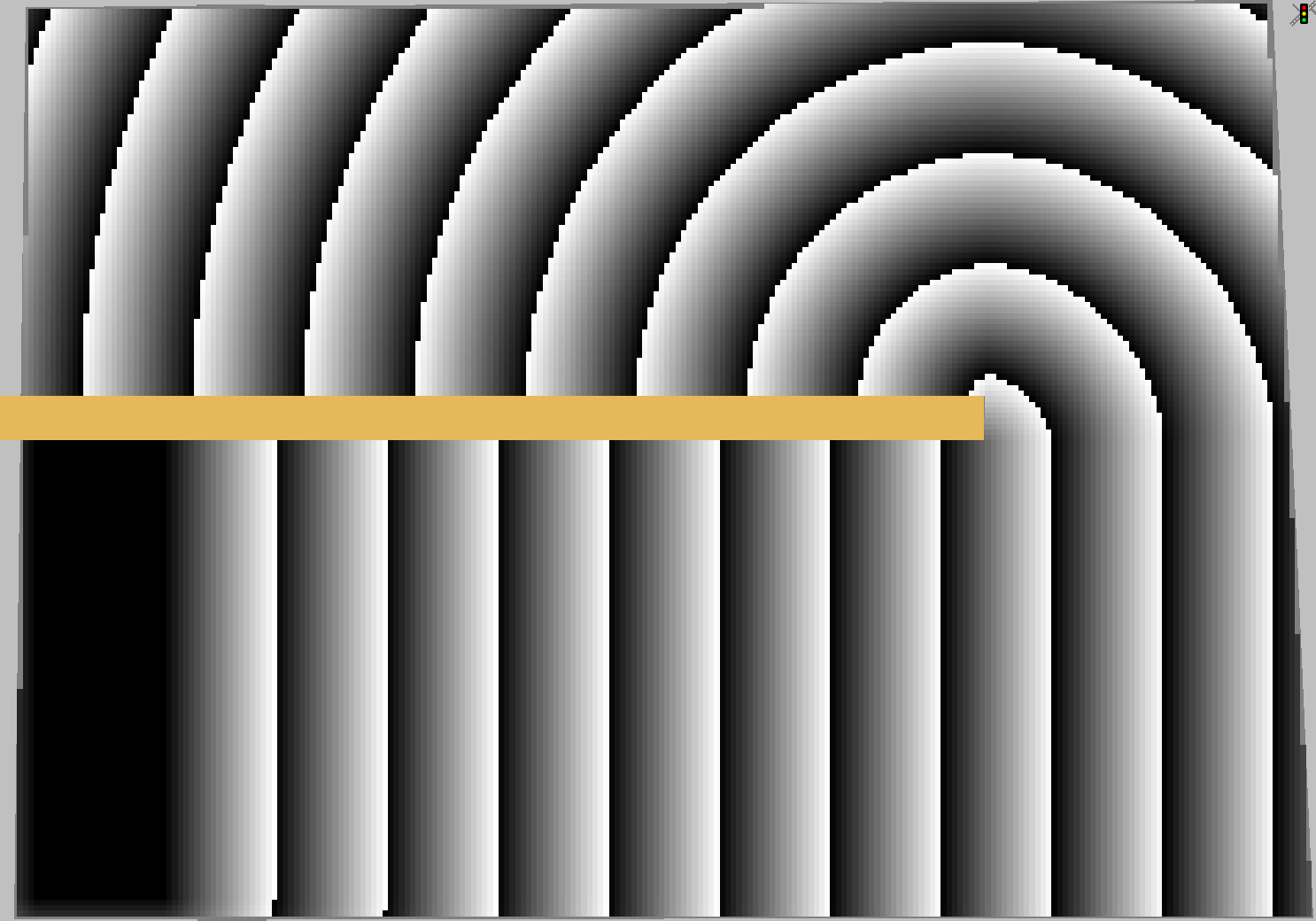} \hspace{12pt}
\includegraphics[width=0.45\textwidth]{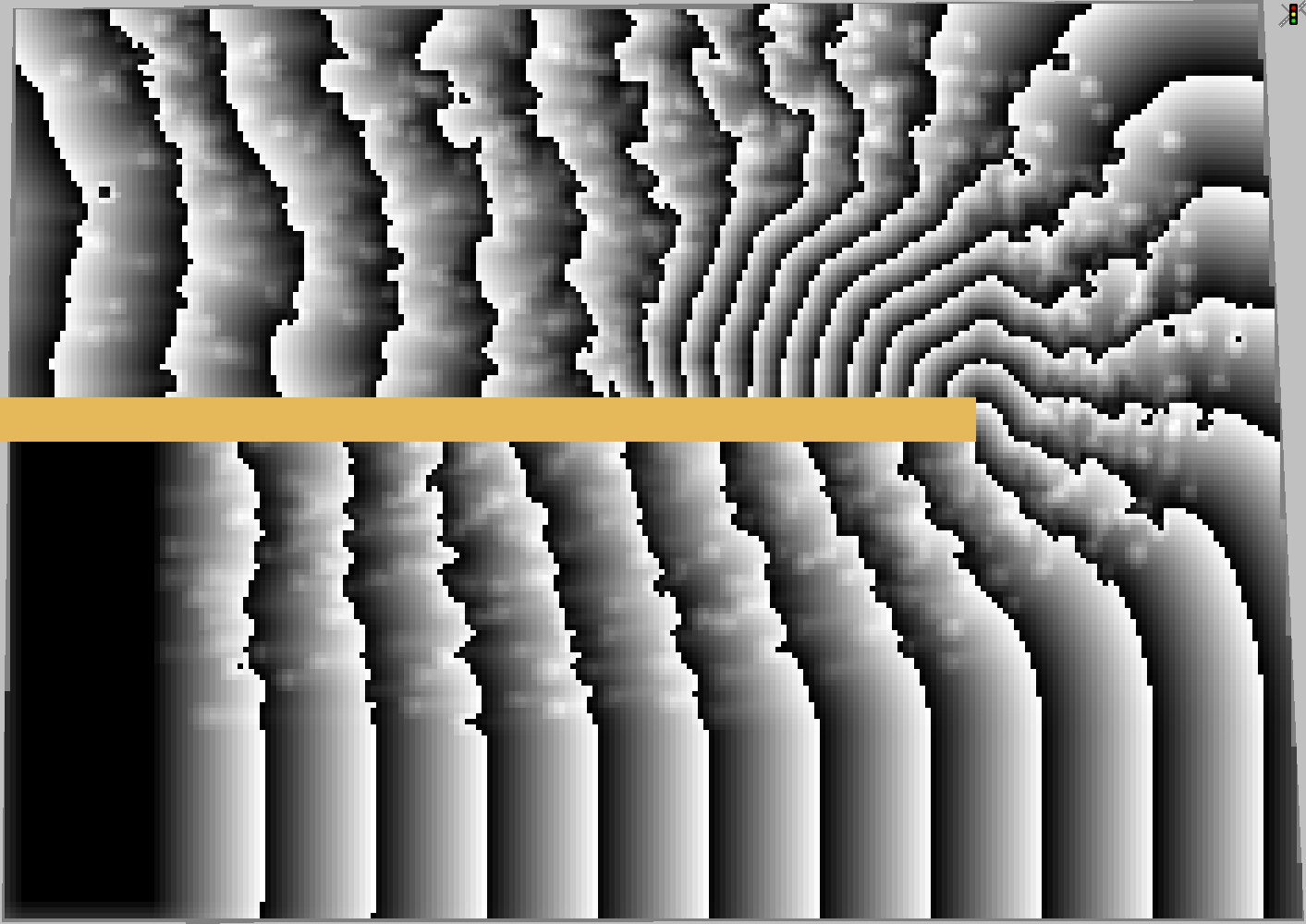}\\
\caption{The static potential (left) and dynamic potential (right) for the last row of figure \ref{fig:U-Turn}.}
\label{fig:U-Turn-Pot}
\end{center}
\end{figure}

\subsection{Station Hall}
\begin{figure}[htbp]
\begin{center}
\includegraphics[width=0.45\textwidth]{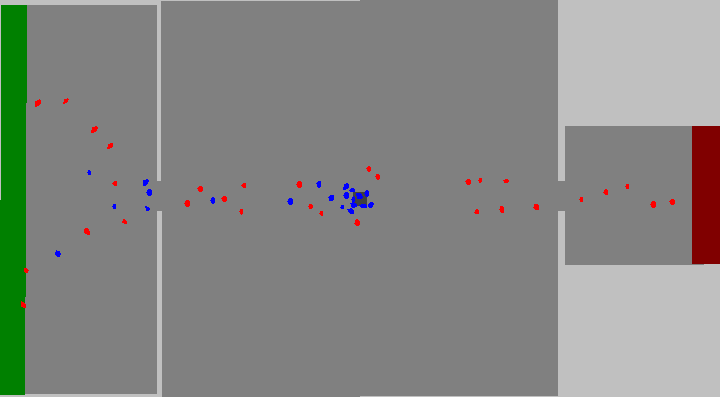} \hspace{12pt}
\includegraphics[width=0.45\textwidth]{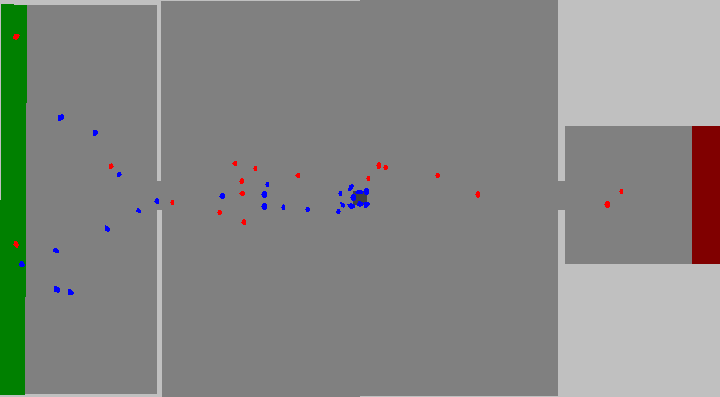}\\ \vspace{6pt}
\includegraphics[width=0.45\textwidth]{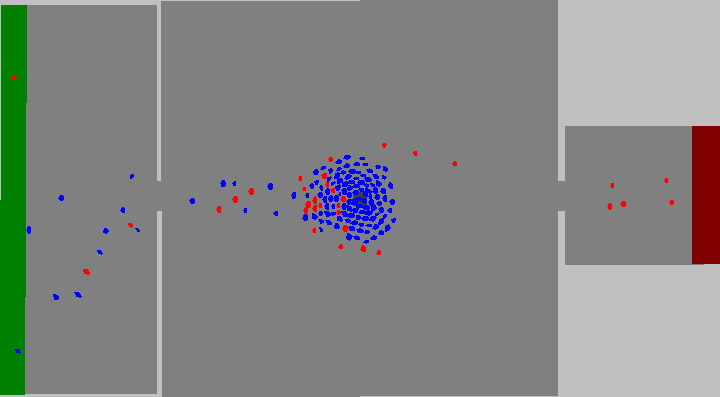} \hspace{12pt}
\includegraphics[width=0.45\textwidth]{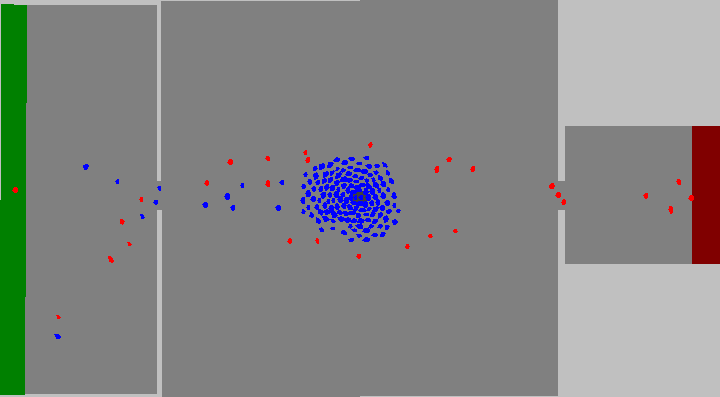}\\ \vspace{6pt}
\includegraphics[width=0.45\textwidth]{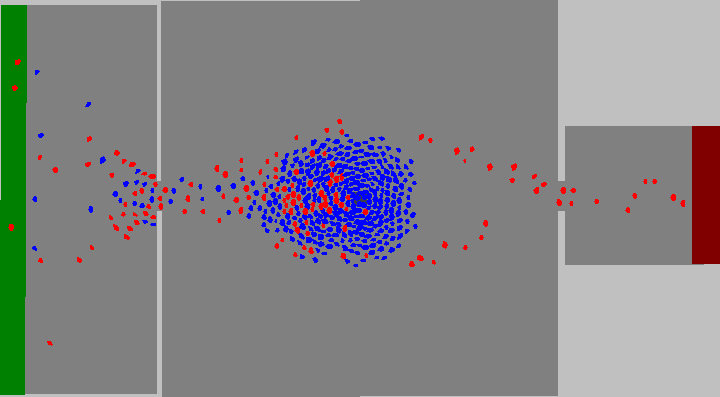} \hspace{12pt}
\includegraphics[width=0.45\textwidth]{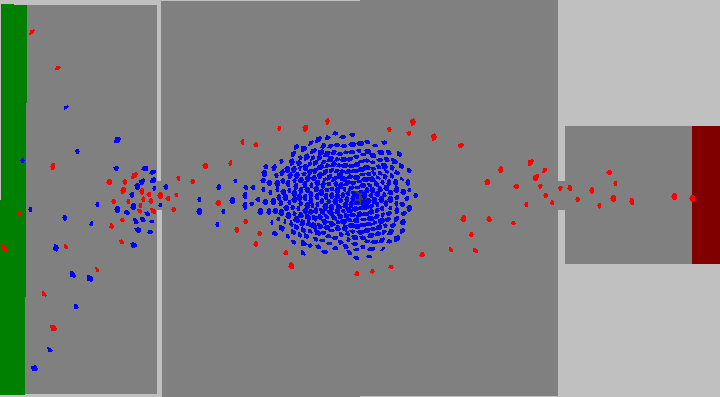}\\ \vspace{6pt}
\includegraphics[width=0.45\textwidth]{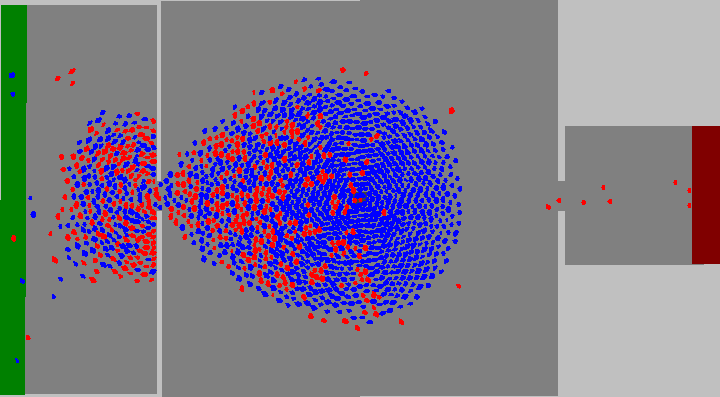} \hspace{12pt}
\includegraphics[width=0.45\textwidth]{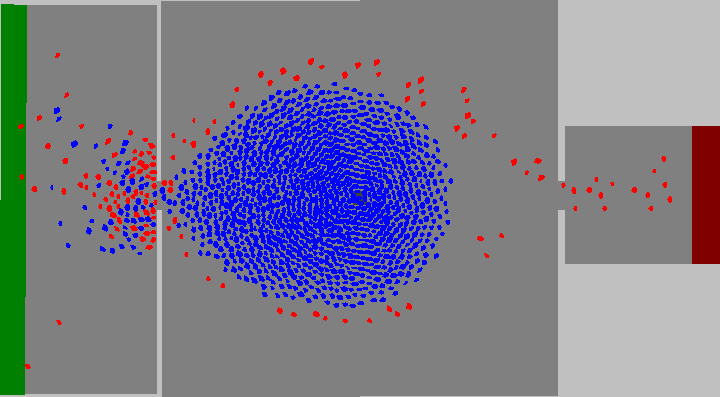}\\ \vspace{6pt}
\caption{Agents are set into the simulation on the green area. The blue agents head for the dark gray area in the middle to dwell there, the red ones want to reach the dark red area. The left column shows the development, if the dynamic potential is not activated, the right column shows what happens, if it is activated. The screen shots were taken after 60, 300, 600, and 1200 simulation seconds.}
\label{fig:stationhall}
\end{center}
\end{figure}

This example demonstrates the efficiency of the method and -- compared to other situations where the method can be helpful -- it is a forgiving example as the results are stable over a wide range of parameters and a number of model variants. The basic idea is that of a station hall, where some people need to hurry to catch their train (travel time is the single-most important movement factor) and others have plenty of time lingering around in groups in the hall. Figure \ref{fig:stationhall} shows an example with only one waiting group (blue agents) that grows over time. The red agents are heading directly to the red area on the right side.

\begin{figure}[htbp]
\begin{center}
\includegraphics[width=\figurewidth]{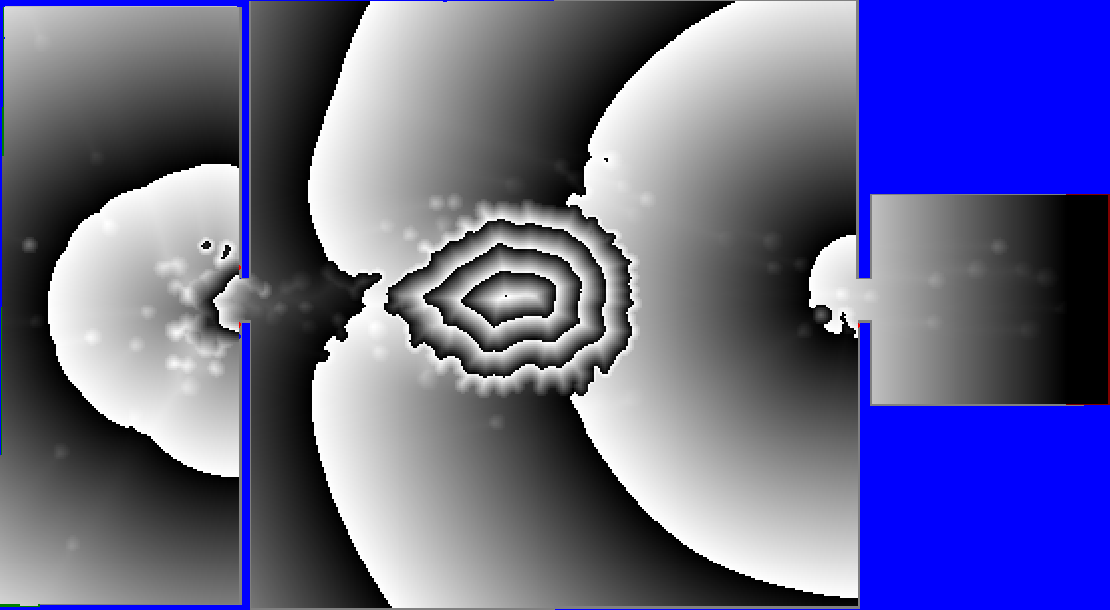}
\caption{The dynamic potential after 600 simulation seconds. The estimated travel time for red agents increases from dark to bright. To increase the contrast the values are once more shown by a modulo.}
\label{fig:pot}
\end{center}
\end{figure}

If the dynamic potential is not activated the red agents get stuck in the group of blue ones as the repulsive forces between agents are not sufficient to make them evade the large group early enough, and those agents that do make it around the group of blue agents are subject to considerable delay. With the dynamic potential the red agents make a detour around the blue group -- no matter how large it grows -- and by avoiding to get stuck in the jam manage to reach the destination in reasonable time, which should be close to the minimal realistic travel time under given circumstances. Figure \ref{fig:pot} shows the dynamic potential after 600 seconds.

The station hall example also shows why the improved behavior cannot be achieved with an increased strength of the repulsive inter-agent forces while keeping the direction of the shortest path as desired direction: the bunch of blue agents is about symmetric where the red agents enter the hall. Therefore the deviating forces from the left and right side would cancel at least partially. Realistically there will never be an exact cancellation along the whole central line toward the group of blue agents, thus the red agents will deviate to one side. Nevertheless, there where they have to deviate (at the entrance) the forces would cancel mutually, while when a red pedestrian is exactly to the side of the blue group and no additional deviation would be needed, the forces from the blue ones would add for further deviation of the red one.

\subsection{Movement around a 90 Degree Corner}
This example is modeled after the geometry of the walking path between the station ``Messe S{\"u}d (Eichkamp)'' and the southern entrance of Berlin's International Congress Center (ICC). This was chosen so that for many readers there is a chance to have been there themselves. It is not exactly a laboratory example, as the corner is rounded and has a bit less than 90 degree. However, the trains set a pulsed demand and with the stairs upper end one can assume that there is a line that by most people is experienced to have equal utility everywhere which means that probably no one on the stairs is thinking of the upcoming corner and moving to either side to improve the position with respect to the corner.

\begin{figure}[htbp]
\begin{center}
\includegraphics[width=0.45\textwidth]{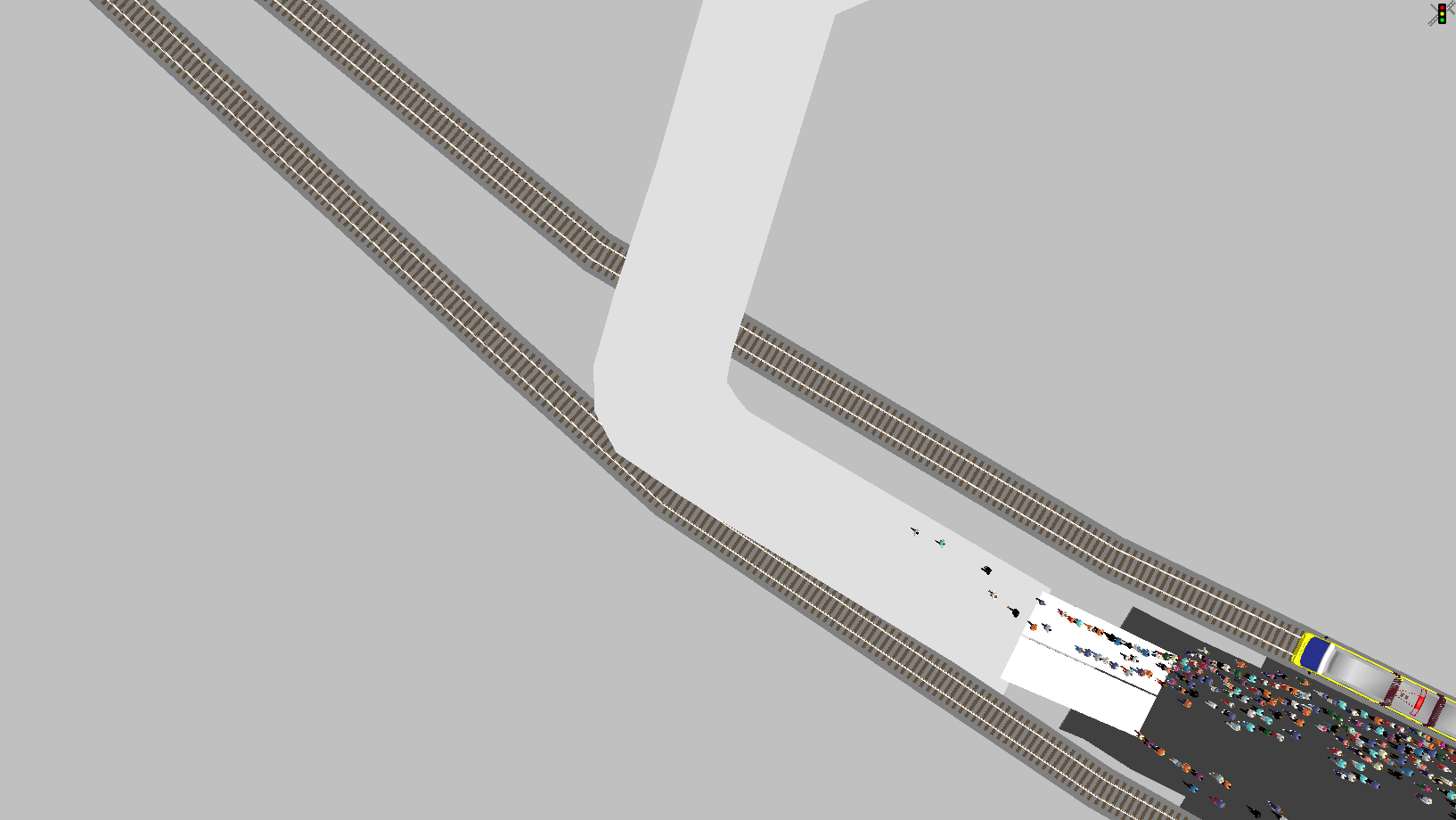} \hspace{12pt}
\includegraphics[width=0.45\textwidth]{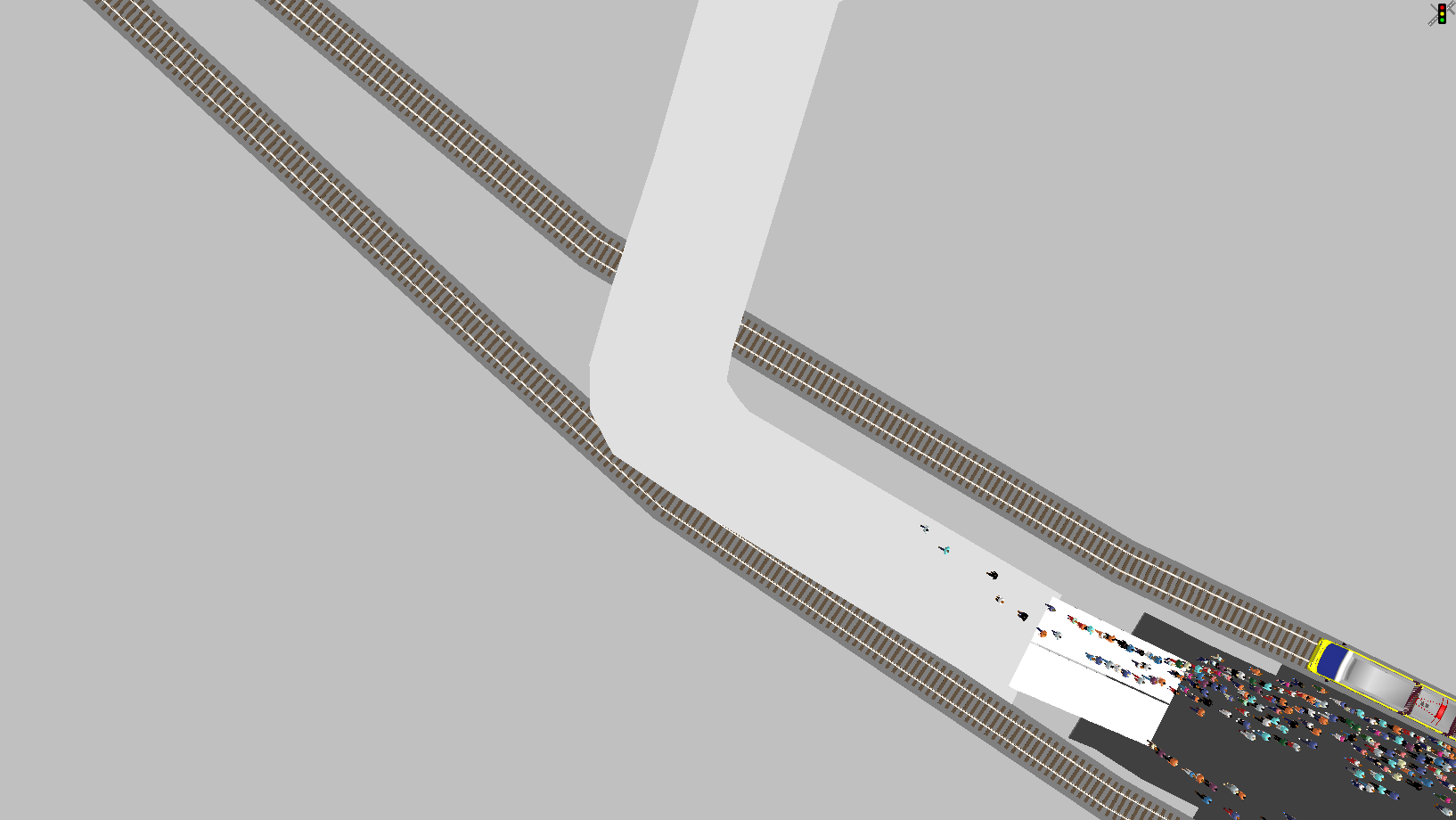}\\ \vspace{12pt}
\includegraphics[width=0.45\textwidth]{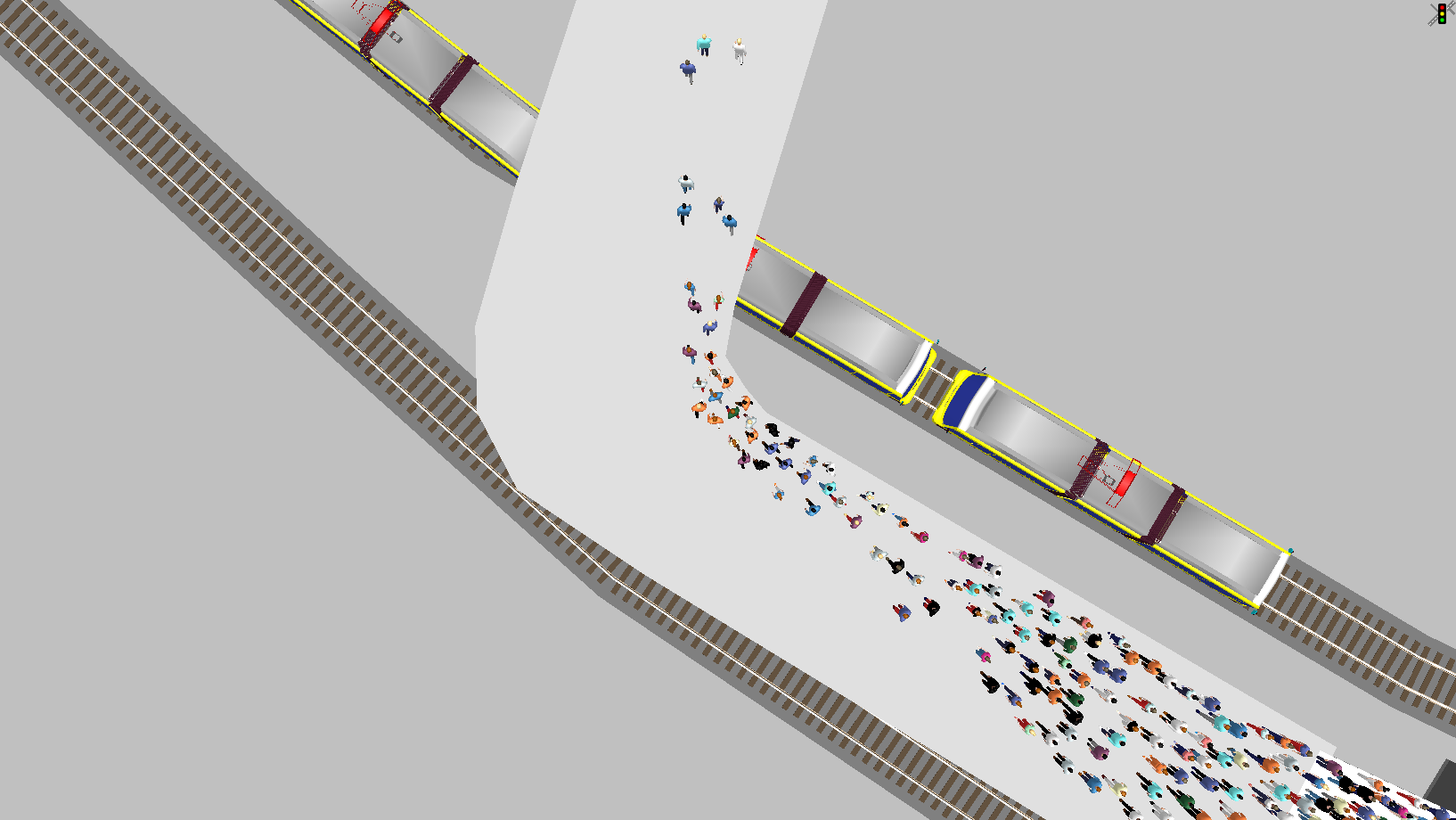} \hspace{12pt}
\includegraphics[width=0.45\textwidth]{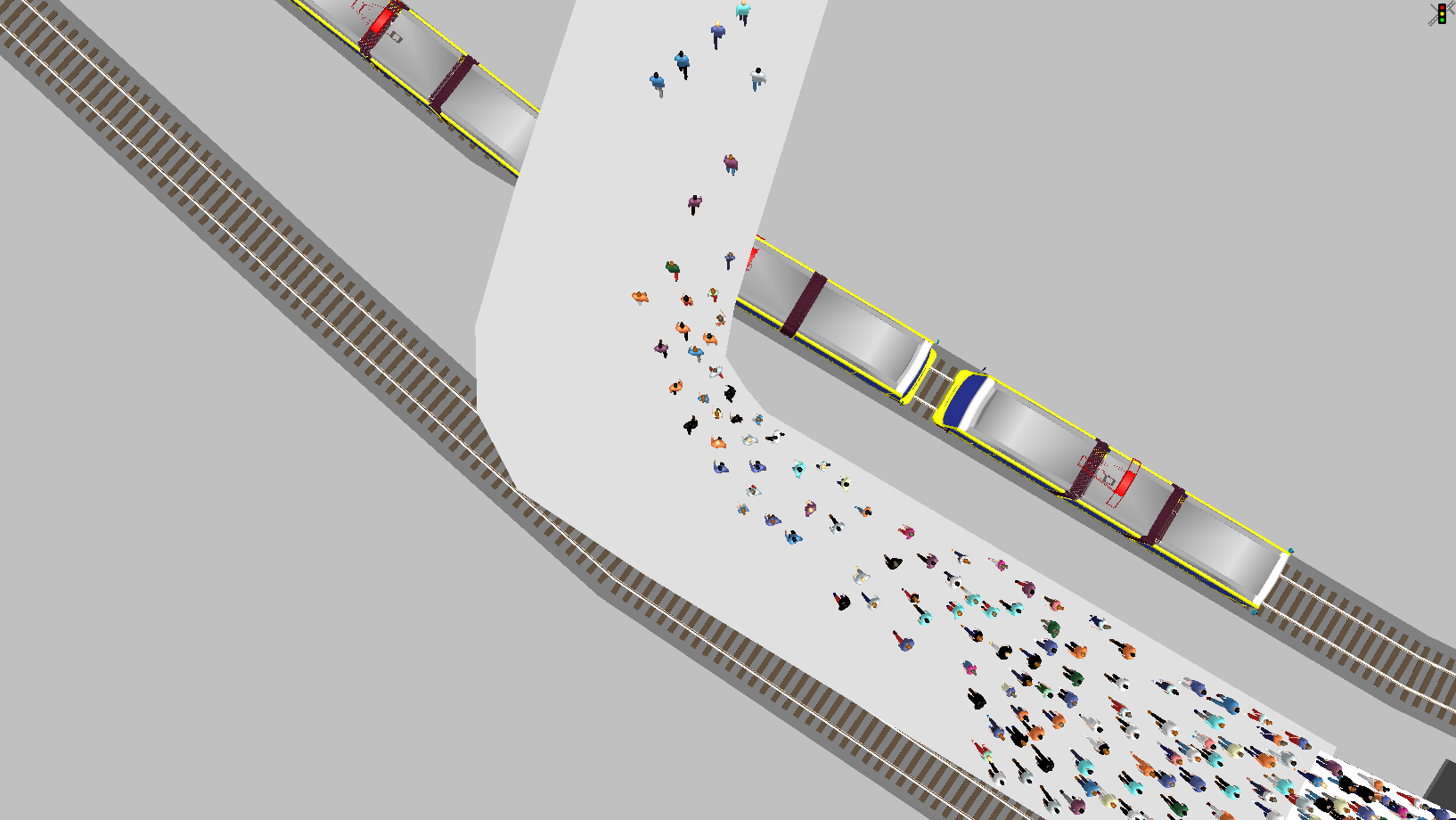}\\ \vspace{12pt}
\includegraphics[width=0.45\textwidth]{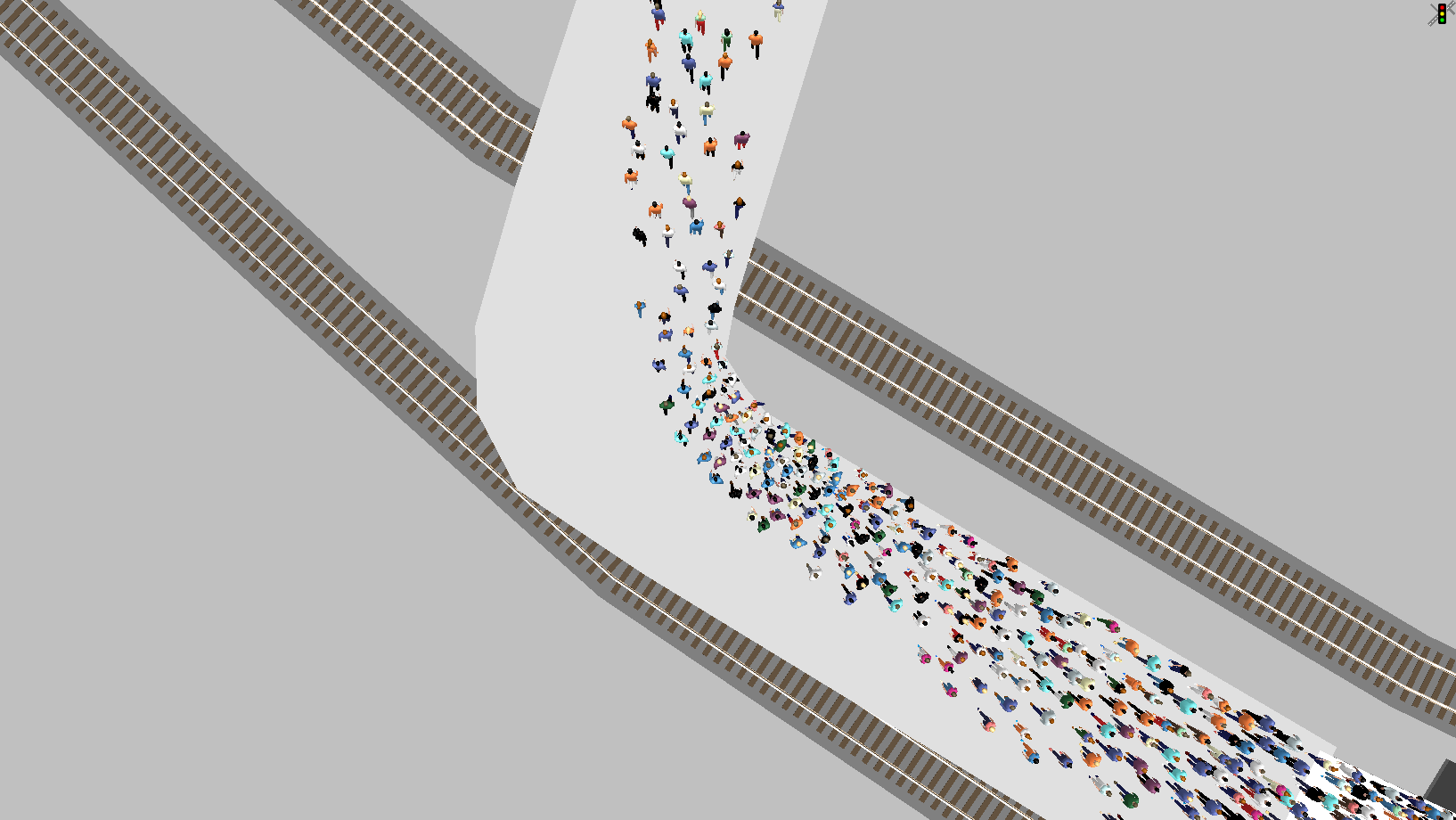} \hspace{12pt}
\includegraphics[width=0.45\textwidth]{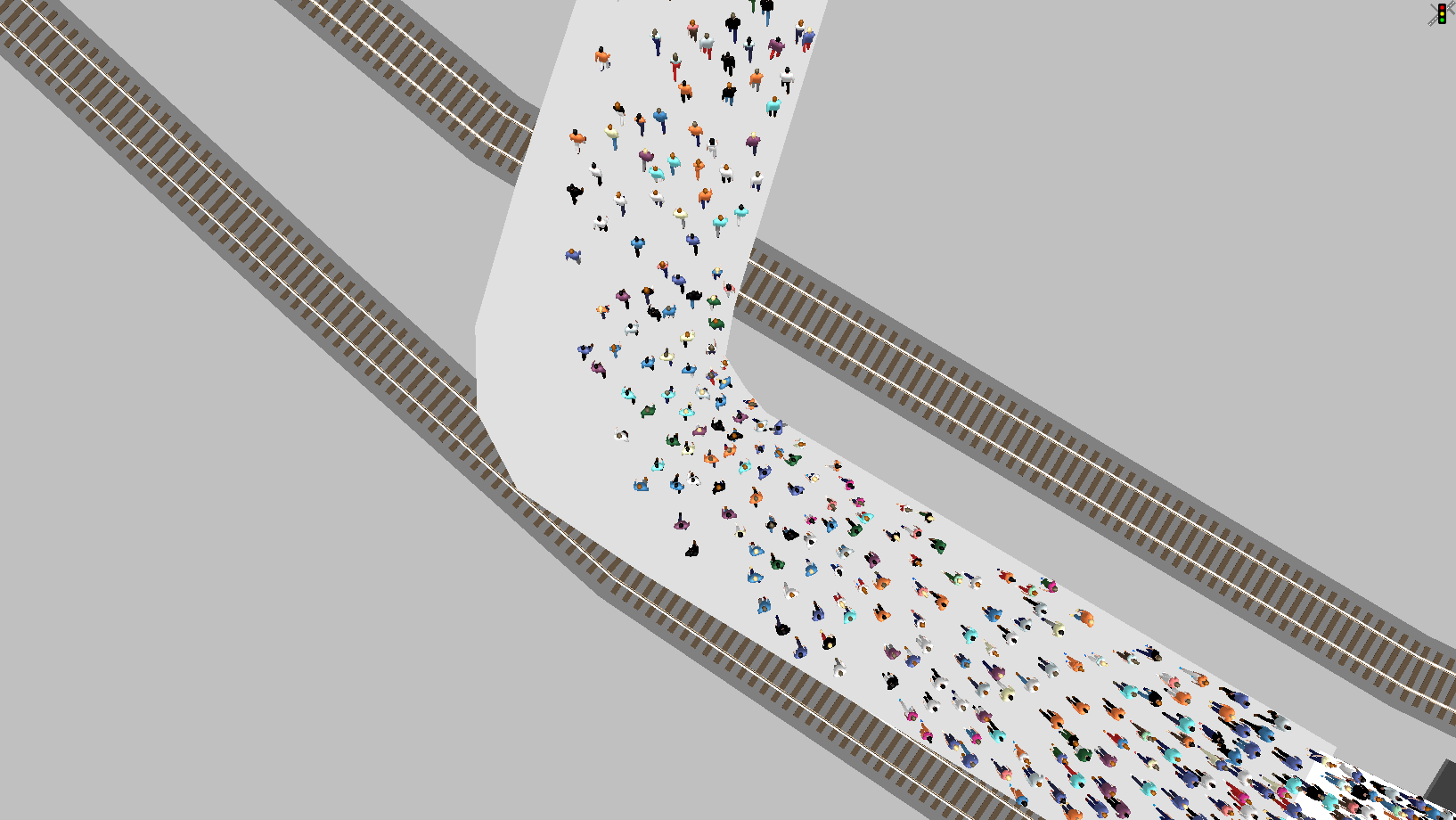}\\ \vspace{12pt}
\includegraphics[width=0.45\textwidth]{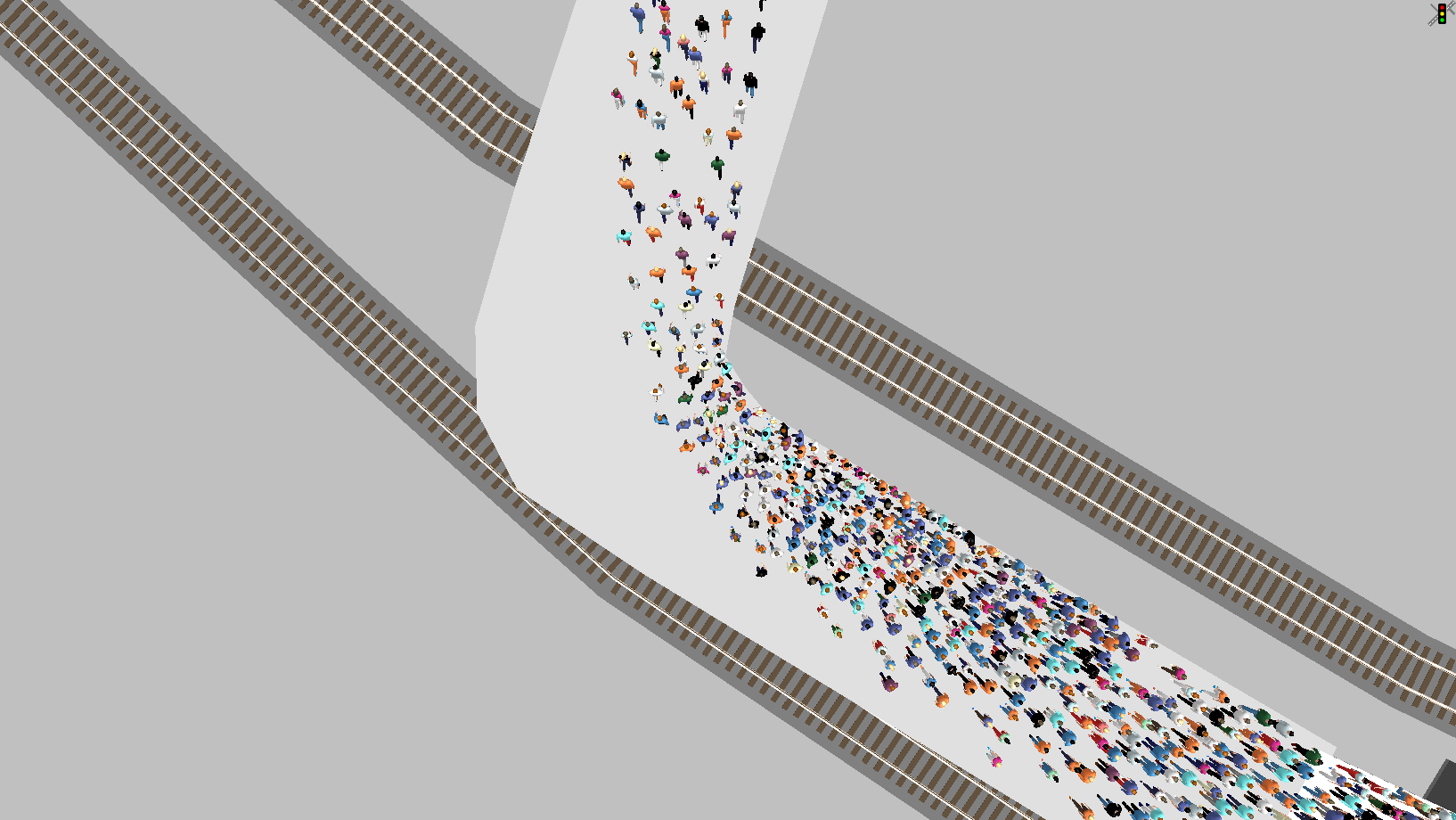} \hspace{12pt}
\includegraphics[width=0.45\textwidth]{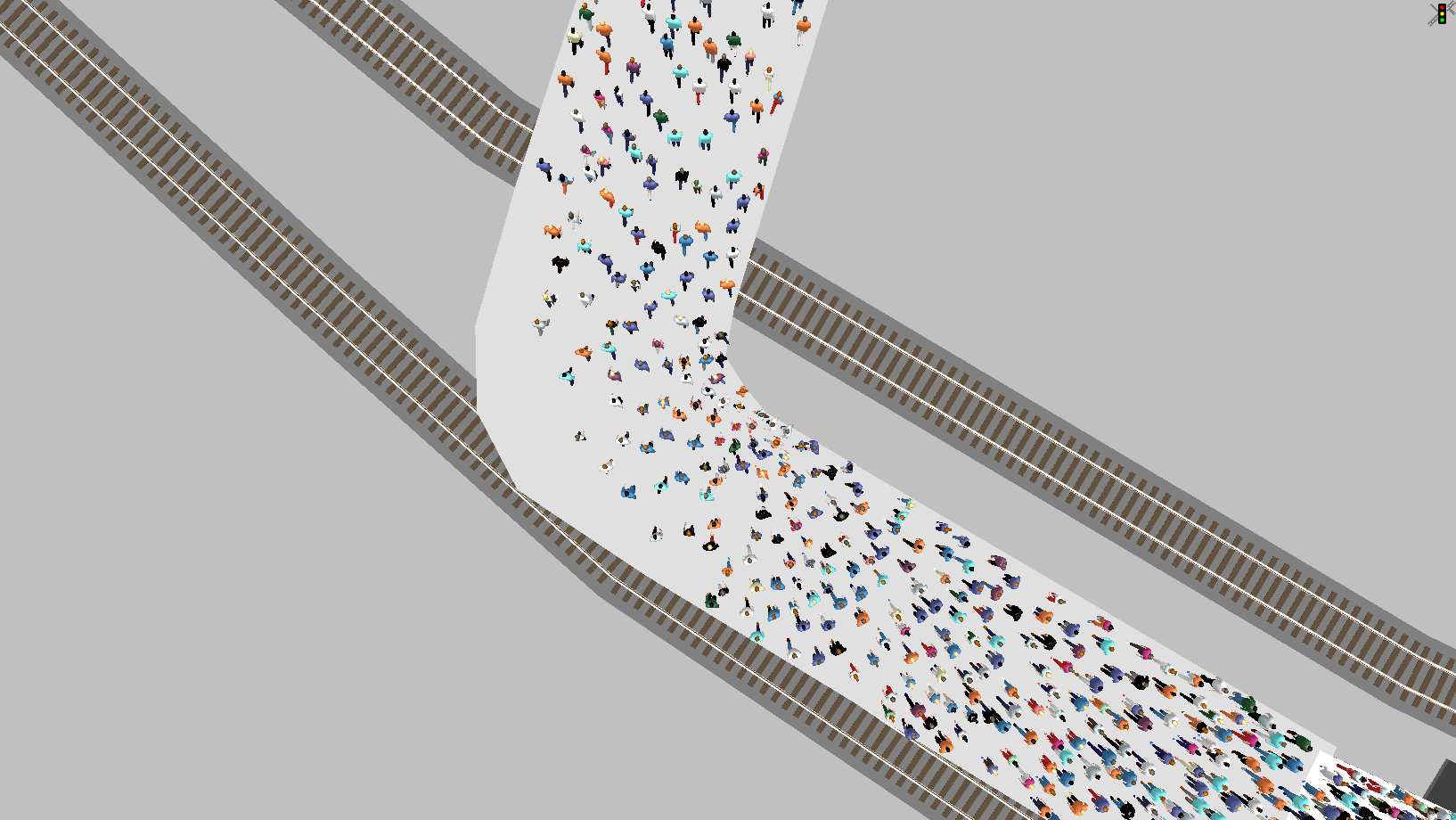}\\ \vspace{12pt}
\includegraphics[width=0.45\textwidth]{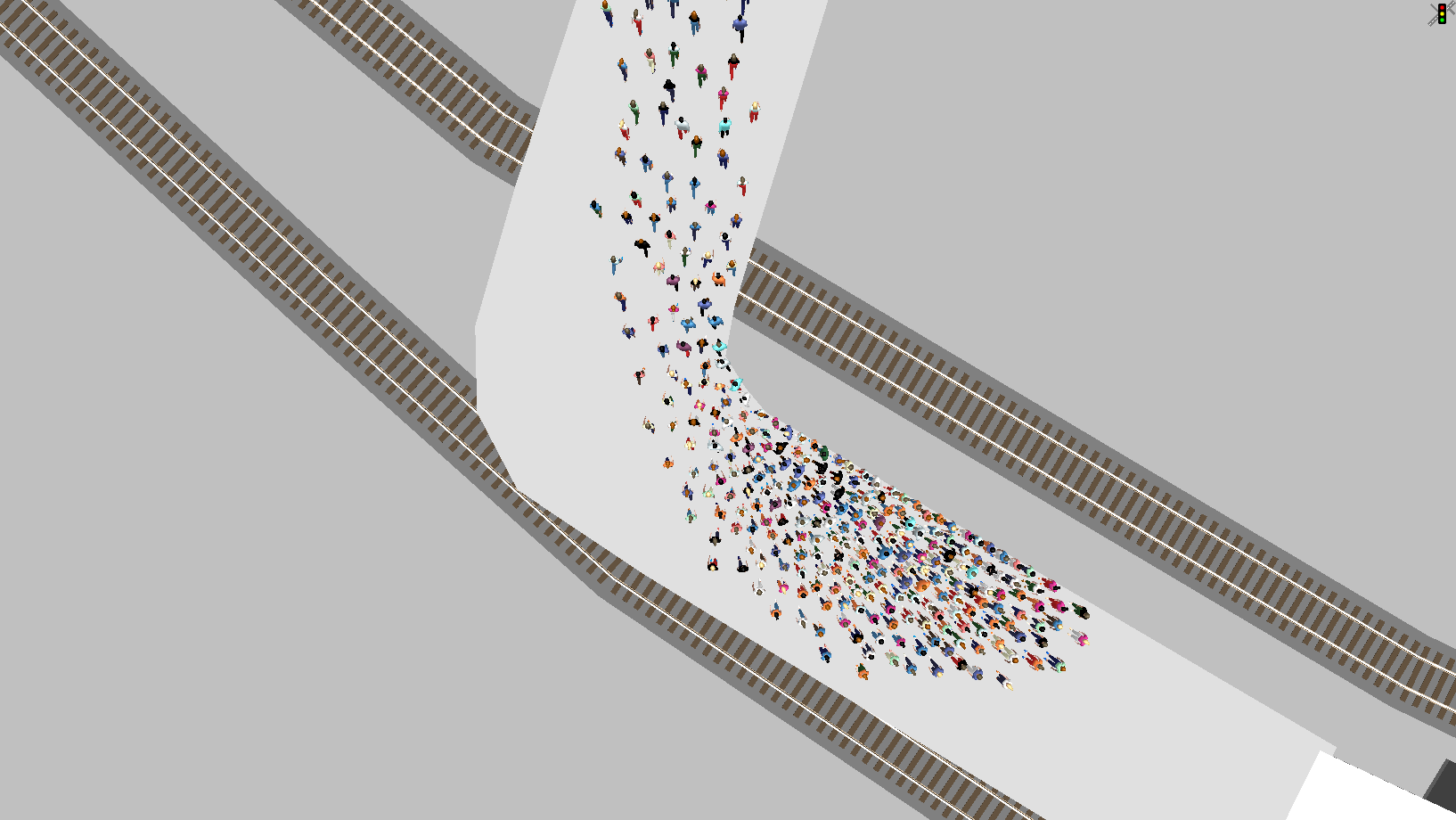} \hspace{12pt}
\includegraphics[width=0.45\textwidth]{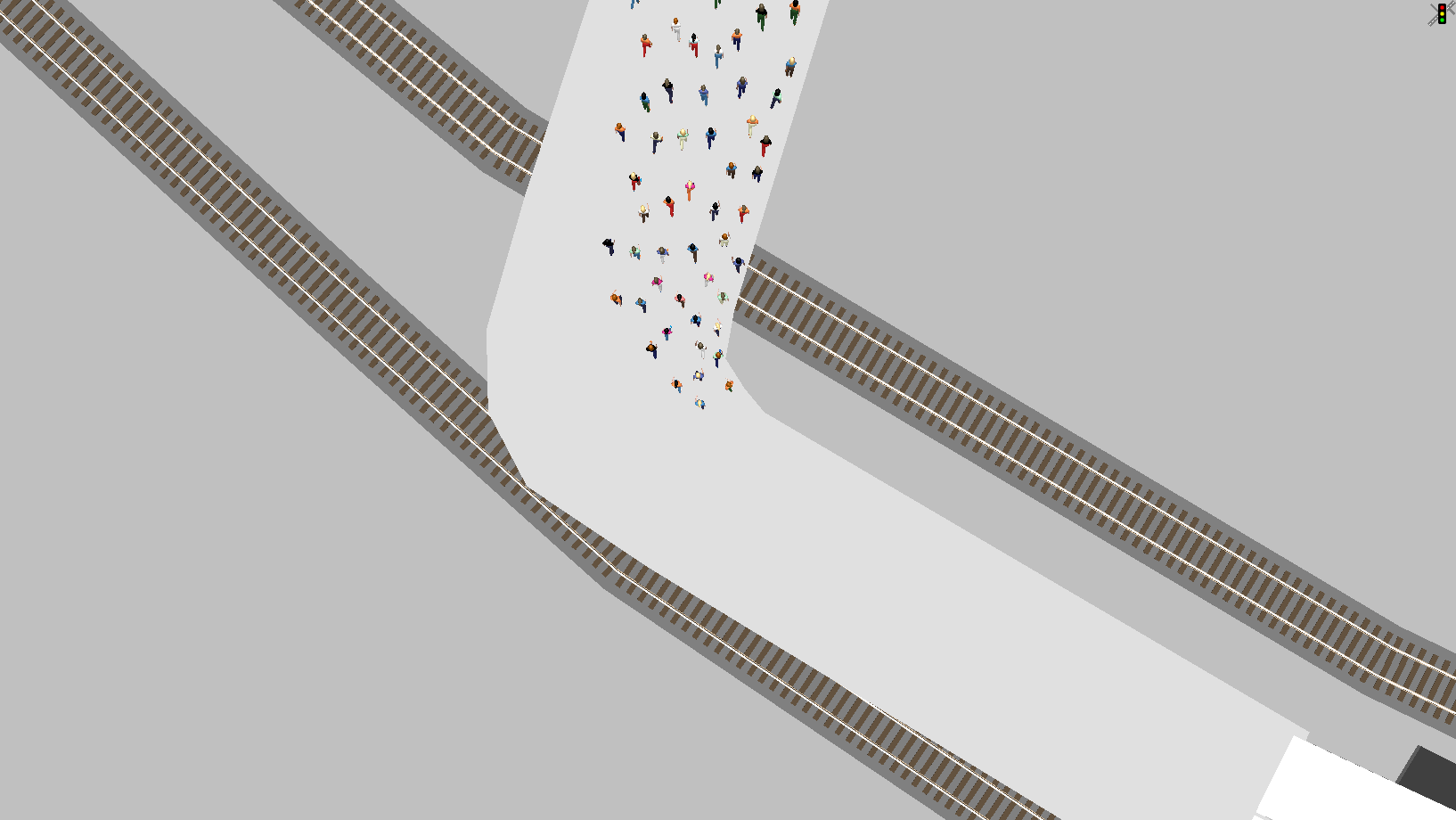}\\ \vspace{12pt}
\caption{Comparison of simulations without (left column) and with (right column) usage of dynamic potential in a scenario modeled after the pathway from Berlin's ``Messe S{\"u}d (Eichkamp)'' station to the southern entrance of the ICC. Between the screenshots there are always 40 seconds except for the last interval, which is 280 seconds.}
\label{fig:ICC} 
\end{center}
\end{figure}

In the simulation it has been assumed that two trains with 800 passengers each arrive at about the same time. On the platform level in both cases (with dynamic potential and with static potential only) the population of agents is identical and the pattern how they leave the trains is  identical as well. This means that the initial conditions are identical. The two scenarios start to diverge when the first agents have climbed the stairs. See figure \ref{fig:ICC}.

For this and the next example compare \cite{Rogsch2010}.

\section{Computation Times}
\begin{figure}[htbp]
\begin{center}
\includegraphics[width=\figurewidth]{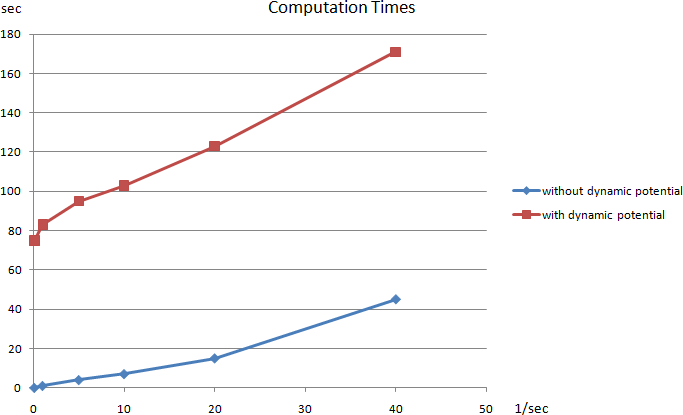}
\caption{Computation time to simulate 100 seconds versus demand in a simple scenario with and without dynamic potential. The demands were 0.1, 1, 5, 10, 20, and 40 pedestrians per second input to the simulation.}
\label{fig:computationtimes} 
\end{center}
\end{figure}

To give an impression of the computational effort we regard an example where pedestrians walk uni-directional from one edge of a 50 m X 50 m square to the other one (parallel to the other two edges). This has been done with various demands from 0.1 to 40 pedestrians per second. Pedestrians are set into the simulation statistically equally distributed over time. The simulation starts with an empty scenario and it is measured how long it takes to simulate from simulation second $t_0=100$ to $t_1=200$. The dynamic potential is recalculated each 0.1 seconds. Figure \ref{fig:computationtimes} shows the results.

\section{Discussion}
\subsection{Open Issues} \label{subsec:OpenIssues}
With the examples of the last section it has been demonstrated that the method proposed in this contribution brings improvements for the simulation of pedestrians in various situations. However, there can be objections from basic theoretical considerations as well as possible variants for elements of the method. These are now discussed which makes this subsection a combination of ``Limitations'' and ``Future Work''.

First and foremost -- if the method is seen as one to approximate the equilibrium of travel times -- it is a non-iterative approach. With regard to an equilibrium it can for fundamental reasons not yield a ``correct'' solution. This is a correct objection, it can only be countered by noticing that a method needs to bring an improvement not perfection to justify its application. A second reply is that it is not yet clarified how close real pedestrian traffic comes to a user-equilibrium. In principle it might be that the agents' behavior as produced by this method has a higher degree of realism than an exact equilibrium. Nevertheless would it be interesting to have a method that is able to produce a user-equilibrium and a system optimum for pedestrian traffic to lay out for example an optimal emergency egress plan. A theoretically profound solution for this issue is not available. A pragmatic approach would be to find parameters $g$ and $h$ or even a different functional form for equation (\ref{eq:f1}) -- maybe using a genetic optimization approach -- that demonstrate in a number of cases that the travel times they estimate are close to the actual travel times (compared after the simulation). 

Second, the impact of an agent's presence and velocity at some spot can have an instantaneous effect on all positions with higher value of the dynamic potential (upstream positions). In this way strictly speaking it is modeled that an agent that is still far away from that spot, but whose path might come close to it, assumes that the situation there will stay like this until it is there. This objection is correct, however in many situations it is not relevant. The impact of a single agent quickly diminishes with distance. The impact on any position more than a few meters aways is marginal and negligible. Only the impact of groups of (jammed) agents is propagated further. However, if there is a jam somewhere it will need some time until it dissolves. It might even be that the system is in a steady-state and the size of the jam remains about constant. Therefore it is justified that a whole group of agents has a longer reaching effect, also affecting agents that would pass by there only later. 

Another issue that calls for an iterative approach is that in equation (\ref{eq:f1}) the scalar product of the agent's velocity with the {\em static} potential is calculated. The idea of this scalar product is to get a numerical value if an agent walks along the main direction of the dynamic potential or if it comes across. For this, however, the scalar product would have to be calculated of the velocity with the {\em dynamic} potential. Yet at this point of the method this is not possible as it is a step in the calculation of the dynamic potential, which therefore is not available yet. The solution would be to calculate the dynamic potential iteratively for a single time step and use the gradient of the dynamic potential of the preceding iteration for the calculation of the next iteration. At the current stage of the implementation of the method this is not done, as it would imply even longer computation times and as the method as it is already has shown to be helpful. If the iterative process will converge to a stable solution for the dynamic potential is another story.

A problem that has shortly been mentioned in the description of the method is that in the current implementation there is only one dynamic potential per destination instead of one per agent. This blurs the effect that equation (\ref{eq:f1}) should depend on the desired speed of the agent on which the dynamic potential is to act. A solution to avoid to have a dynamic potential for each agent is to have dynamic potentials for ranges of desired speeds. This would still increase the computation time by a factor of two to ten, but it would be manageable. Yet there is another reason why one would like to have one potential per affected agent: only in this way could an agent be affected by a potential on which it has not had an impact on before. In the method as it is proposed each agent has an impact on the dynamic potential and later is influenced by that same potential. This implies that the agent in its movement is affected by its own presence and velocity (compare section 6.1 of \cite{Kretz2009a}). This can lead to undesired side-effects, which can only be guaranteed to be controlled if parameter $g$ is kept to moderate values of approximately $g<3.0$ m/s (with grid spacing constant 20 cm and above mentioned radius of influence).

It is possible to use values of $g$ that are larger, if one takes care that the undesired side-effects are suppressed. One way to do this is to use a mix of the direction of the shortest path and the direction of the estimated least travel time as direction of the desired velocity:
\begin{equation}
\vec{v}_0 = v_0\frac{p\frac{-\nabla T}{|\nabla T|} + (1-p)\frac{-\nabla S}{|\nabla S|}}{\left|p\frac{-\nabla T}{|\nabla T|} + (1-p)\frac{-\nabla S}{|\nabla S|}\right|} \label{eq:transgression}
\end{equation}
with $0\leq p \leq 1$. This allows to continuously switch from static to dynamic potential. Another method is to forbid that for a particular agent the angle between the desired velocity and the direction of the shortest path changes too much within a time step, i.e. if the angle between $\nabla T$ and $\nabla S$ increases too much from one simulation time step to the next, not $-\nabla T$ is used as direction of the desired velocity, but a direction that is closer to $-\nabla S$.

In the method as proposed each agent individually has a roughly circle-shaped impact on the field of $f$. From this follows in the field of $f$ many steps from $1/f=1$ (outside the circles) to a larger value (inside a circle). For the movement of the agents this can result in small but frequent changes in the velocity. Smoother movement might be received if first a density field is calculated for the whole walkable area and then from these densities $f$ is calculated. As long as the density field consists of Voronoi cells \cite{Steffen2010b,Liddle2011} the individual and locatable information of walking velocity remains preserved, if the density is calculated in the way of a probability density \cite{Johansson2008,Hanebeck2008} one could expect even smoother movements, but runs into trouble with locating the velocities.

It has already been stated that equation (\ref{eq:f1}) is not strictly derived, but rather chosen by plausibility. It is not the simplest equation one can think of, this would be $f=\bar{f}<1$ for all grid points marked as occupied. Equation (\ref{eq:f1}) is a rather simple one under the constraint that one wants to consider the velocity of the agent at that spot and that one wants to be able to calibrate both: the dependence of the velocity (via $h$) and the general strength of the impact of an agent (via $g$). From this follows that equation (\ref{eq:f1}) is not guaranteed to be the best choice for this purpose, but it suggests itself as a starting point. An alternative idea would be to make use of a common capacity restraint function from macroscopic transportation planning instead \cite{VISUM2010}. However, this is problematic as capacity restraint functions use the capacity of a link as explicit input. As pedestrian walking infrastructure can be arbitrarily complicated it is not trivial in general to give a capacity although not always impossible \cite{Lew2009,Schomborg2011,Hamacher2011a,Hamacher2011b,Kneidl2011}.

\subsection{Additional Applications}
It should shortly be mentioned here that the method as proposed can perfectly be used to model a few other phenomenons.

The repulsive effect of walls and corners (beyond preventing agents and walls to overlap) can be modeled by assigning a value $f<1$ to grid points, which are close to walls.

Areas with variant average movement speed (a conveyor belt or a sand beach) can directly be taken into account by setting $f=v_{variant}/v_{normal} f_0$ (with $f_0$ applying outside the special area). This will attract or repulse agents in the surrounding to detour to use that area or to avoid it. Note: This does not cause agents on that area to change their desired speed automatically, this needs to be defined in the simulation scenario separately.

Areas like bike or vehicle lanes that are avoided by pedestrians if the density on the walkway is sufficiently low, but which do not cause a change in walking speed, can nevertheless be modeled in the same way. The value of $f$ for these needs to be set such that inundation of pedestrians to the lanes begins at an empirically verified density on the walkway \cite{Lew2009}. $f$ is then more of a gauge parameter whose value cannot be interpreted directly, but depends on the other settings of the scenario. Needless to say that it is also possible to model specifically preferred areas (e.g. a roofed path on a rainy or hot day) just in the same manner.

\subsection{Choosing the Values of $g$ and $h$} \label{subsec:gh}
Parameter $g$ has the role of setting the overall strength of the dynamic potential. If $g=0$ then a static potential is calculated that gives the distances to the destination. If $g \rightarrow \infty$ then the agents are handled as if they were static obstacles. The experience with the method has been that $0 < g < 1$ leads to effects which are barely noticeable\footnote{All these statements are made on the background of a spacing of 20 cm between the grid points.}, i.e. agents walk more or less on the shortest path. A good value appears to be $g=1.5$ where a positive effect is visible, while undesired side-effects hold off. For difficult situations as a u-turn or situations with discrete choices (ticket gates in a bending) a value of $g=1.5$ might be too weak and increasing it to about $g=3.0$ is an option. Above that value the side-effects become annoyingly visible.

A large value of $g$ can also induce a problem in combination with the fact that the gradient calculation is a difference and not a differential quotient. If $g$ is large then it may happen that the dynamic potential flows into the area occupied by the agent from the front and from the back. The center will always be reached from the front, but if the grid point used for gradient calculation which lies to the back is reached from the back, the gradient might reverse its direction and by that result in entirely unrealistic directions of the desired velocity. This problem is reduced, when the grid point spacing is reduced and it would vanish, if there was one potential per agent and each agent would not have an impact on the potential which it is influenced by.

Parameter $h$ determines how much the velocity of an agent is taken into account, when its impact on the field of $f$ is calculated. With $h=0$ there is no impact, with $h=1$ there is a full impact as so far as an agent has no impact on the field of $f$ when it moves with $v_0$ exactly in the direction of the static potential. It does not cause problems to set $h>1$, although it probably does also not help. The experience with $h$ so far is that good values lie in $0.6\leq h\leq 0.8$. At the upper end agents appear to mainly react on existing jams, while at the lower end of the interval agents seem to expect that a dense but still moving crowd soon will form a jam. The latter case gives the impression of stronger foresight but also introduces more disturbance while the crowd is moving.

\subsection{Existing Work}
This section is to sharpen the profile and intentions of this contribution by distinguishing it from related existing work.

In \cite{Hoogendoorn2004} Hoogendoorn and Bovy introduced a very profound theory of pedestrian dynamics based on utility maximization. The (dis)utilities affecting the route choice are travel time, kinetic energy (i.e. walking speed), nearby obstacles, density as determinant of unwanted contacts, and the stimulation of the environment. Hoogendoorn and Bovy take higher walking speeds as disutilities into account just as higher travel times. A time-pressure co{\"e}fficient controls exogenously, if a pedestrian rather accepts the disutility of higher walking speeds or higher travel times. Additionally the desired walking speed is affected by the absolute value of the gradient of the utility field: if the utility to be at another nearby spot is not much different from the utility of the current spot then the agent walks slower. In our method the absolute value of the gradient of the field of remaining walking times does not have an impact on the desired walking speed. Only the normalized direction is used for the desired walking direction. This has three reasons: first off, the desired speed is a more manifest parameter than a time-pressure co{\"e}fficient. And second, it needs to be shown, if pedestrians really slow down, if utility cannot be increased much now, as long as there is a chance for a steeper increase in utility later. 

Treuille et al \cite{Treuille2006} also have introduced a model with (dis-)utility as determinant of the movement. There however it is not stated which value of $f$ is used at occupied spots. As the authors do not discuss this issue, one can assume that it is globally a constant value, maybe even $f=0$, i.e. that occupied spots are treated as if they were walls. 

The work of Hughes \cite{Hughes2002,Hughes2003} is more similar to this contribution as it closely links travel time and density and focuses on travel time as the determining factor. However, it is a macroscopic first-order theory and as such only loosely related to our intention to enhance a microscopic simulation of pedestrian dynamics.

There is other preceding work \cite{Kretz2009a,Kretz2009c,Kretz2010c,Kretz2010a,Kretz2010b} which adopted a very similar approach for a discrete-space (``cellular automata type'') model but used a different method to calculate the direction determining fields, the reason for this being a very high computational efficiency of that method. As a downside the field suffers from deviations from Euclidean metric, even if there is not a single agent present. Anyhow it is possible to transfer the insight gained in these works to the usage of the method with the Social Force Model.

In the field of robotics a related work has been introduced by Shopf et al. \cite{Shopf2008}. They have used the same Eikonal Equation solver and implemented it for execution on a GPU. Differences to our approach are that the model for agent dynamics is a different one which was formulated for robot path planning \cite{Fiorini1998}. As the trajectories of robots do not need to resemble those of humans, this is not sufficient for our task. For example will robots moving according to this model only decelerate if the time to collision is less than the time step  of the simulation (the inverse of the number of simulation steps per second). This resembles to the Nagel-Schreckenberg-Modell with $p=0$ \cite{Nagel1992}. With such a computationally cheap movement model and the Eikonal Equation solver's computational complexity being dependent only on area size and not agent number, the authors manage to simulate 65,000 agents in real time. Under given circumstances this high computation speed might only be possible by having dense crowds (compare Bleiweiss \cite{Bleiweiss2009}). At the same time a real-time simulation capability of 180,000 agents in a model without Eikonal Solver \cite{Kretz2010b} shows that inevitably the method will have a major share of the computation time, no matter how efficiently computed and no matter what model it is used with.

Kirik et al have considered and modeled the issue of quickest path movement in a simulation of pedestrians in a cellular automata-derived model \cite{Kirik2009,Kirik2011}.

The FMM was used in at least two more models of pedestrian dynamics to calculate a static potential \cite{Venel2010,Hartmann2010}.

Dressler et al \cite{Dressler2010} approached the problem of quickest paths in microscopic simulations of pedestrians by relating route choice to previously calculated network flows.

L{\"a}mmel et al \cite{Lammel2010} have approached the problem of quickest paths for pedestrians with a network-based micro simulation.

Guy et al \cite{Guy2010} based pedestrian navigation not on a principle of least remaining travel time, but on least effort.

Ond{\v{r}}ej et al have published a model, where for agent navigation the vision process is directly modeled \cite{Ondrej2010}. In this aspect the model is similar to the very recent Social Force Model update by Moussa{\"i}d et al \cite{Moussaid2011}, however, Ond{\v{r}}ej et al do not base their model on forces.

Kemloh et al have introduced an event-based method that allows local as well as global quickest path egress strategies \cite{Kemloh2011}.

\section{Summary and Conclusions }
In this contribution we have introduced a non-iterative method that estimates the direction of the quickest path for the movement of pedestrians in a microscopic simulation. The method and therefore the calculated direction is continuous with regard to space. We have used the results of the method as input for the Social Force Model of pedestrian dynamics and presented the effect in four different examples. All examples have in common that for many agents the direction of the shortest and the direction of the quickest path differ considerably for relevant time spans. We think it is clearly visible already from the still images in this contribution that calculating the direction of the desired velocity along the quickest path according to the new method gives better results. Having in this way demonstrated the benefit of the method we engaged in a discussion of limitations and possible variants and improvements.

\section{Supplemental Material}
The illustrating example of section \ref{sec:Method}, the examples of section \ref{sec:Examples} plus some additional examples are available as animation at \url{http://www.youtube.com/watch?v=8SmRBTJ-jeU}. The pedestrian simulation model and the extension proposed in this contribution have been implemented in VISSIM \cite{VISSIM2010,Fellendorf2010} and are available for download as a trial version including all functionality relevant for this paper.

Numerical Values for Parameters and Properties of the Simulation: We have found the following values useful and feasible for typical applications:

\begin{itemize}
\item Time step of the pedestrian simulation (Social Force Model): 0.05 seconds.
\item Recalculation time interval for the dynamic potential: 0.1 seconds.
\item Lattice spacing of map of distances as well as dynamic potential (map of estimated remaining travel times): 15 to 20 cm.
\item Parameter g: 1.0 to 2.5, if in equation (\ref{eq:transgression}) parameter $p=1$, else $g$ can be larger.
\item Parameter h: 0.0 to 2.0.
\end{itemize}

%
\nocite{_PED2003,_ACRI2006,_PED2008,_TGF2009,_Meyers2009,_PED2010}
\bibliography{eikonal_sfm}

\bibliographystyle{utphys2011b}

\end{document}